\documentclass[12pt]{article}
\usepackage[pdftex]{graphicx}     
\usepackage{enumerate}
\usepackage{natbib}
\usepackage{url} 
\usepackage{amsmath,amsfonts,enumerate,verbatim,booktabs,amssymb}
\usepackage{algorithm}
\usepackage{microtype}
\usepackage{xspace}
\usepackage{mathtools}
\usepackage[normalem]{ulem} 
\usepackage{times}
\usepackage[dvipsnames]{xcolor}
\usepackage{pdflscape}
\usepackage{scalerel}

\usepackage{xr}
\usepackage[english]{babel}
\graphicspath{{graphics/}}

\usepackage[figuresright]{rotating}

\renewcommand{\vec}[1]{\boldsymbol{#1}}
\newcommand{\lambdavec}{\vec\lambda}
\newcommand{\thetavec}{\vec\theta}

\newcommand{\sigmavec}{\vec\sigma}

\newcommand{\xvec}{\vec{x}}

\newcommand{\vecq}{\vec{q}}
\newcommand{\indep}{\overset{indep}{\sim }}
\newcommand{\mysim}{\overset{\phantom{indep}}{\sim }}

\newcommand\bigfrac[2]{\frac{\displaystyle{#1}}{\displaystyle{#2}}}
\newcommand{\Ga}{\text{Ga}}

\newcommand{\LN}{\text{LN}}
\newcommand{\E}{\text{E}}
\newcommand{\Bern}{\text{Bern}}
\newcommand{\Po}{\text{Po}}
\newcommand{\D}{\mathcal{D}}

\title{On Bayesian inference for the Extended Plackett-Luce model}

\author{Stephen R. Johnson , Daniel A. Henderson, Richard J. Boys}
\date{\small School of Mathematics, Statistics and Physics, Newcastle University, UK} 

\begin{document}

\def\spacingset#1{\renewcommand{\baselinestretch}%
{#1}\small\normalsize} \spacingset{1}

\maketitle 

\begin{abstract}
The analysis of rank ordered data has a long history in the statistical literature across a diverse range of applications.
In this paper we consider the Extended Plackett-Luce model that induces a flexible (discrete) distribution over permutations.
The parameter space of this
distribution is a combination of potentially high-dimensional discrete
and continuous components and this presents challenges for parameter
interpretability and also posterior computation.
Particular emphasis is placed on the interpretation of the parameters in terms of observable quantities and we propose a general framework for preserving the mode of the prior predictive distribution.
Posterior sampling is achieved using an effective simulation based approach that does not require imposing restrictions on the parameter space.
Working in the Bayesian framework permits a natural representation of the posterior predictive distribution and we draw on this distribution to address the rank aggregation problem and also to identify potential lack of model fit. The flexibility of the Extended Plackett-Luce model along with the effectiveness of the proposed sampling scheme are demonstrated using several simulation studies and real data examples.

\end{abstract}

\noindent%

\spacingset{1.45} 

\section{Introduction}
\label{sec:introduction}

Rank ordered data arise in many areas of application and a wide range
of models have been proposed for their analysis; for an overview see
\cite{Marden95} and \cite{alvo2014statistical}.  In this paper we
focus on the Extended Plackett-Luce (EPL) model proposed by
\cite{MollicaT14}; this model is a flexible generalisation of the
popular Plackett-Luce model
\citep{luce1959indicidual,plackett1975analysis} for permutations.
In the Plackett-Luce model, entity $k\in\mathcal{K} = \{1,\dots,K\}$ is
assigned parameter $\lambda_k>0$, and the
probability of observing the ordering $\vec{x}=(x_1,x_2,\ldots,x_K)'$ (where $x_j$ denotes the entity
ranked in position $j$) given the entity
parameters  $\vec{\lambda}=(\lambda_1,\ldots,\lambda_K)'$ is 
\begin{equation}
\Pr(\vec{X} = \vec{x} |\lambdavec) = \prod_{j=1}^{K}
\frac{\lambda_{{x_{j}}}}{\sum_{m=j}^{K} \lambda_{{x_{m}}}}. 
\label{eqn:spl_prob}
\end{equation}
We refer to \eqref{eqn:spl_prob} as the \textit{standard} 
Plackett-Luce probability. This probability is constructed via the so-called
``\textit{forward ranking process}'' \citep{MollicaT14}, that is, it
is assumed that a rank ordering is formed by allocating entities from
most to least preferred.  This is a rather strong assumption.  It is
easy to imagine a scenario where an individual ranker might assign
entities to positions/ranks in an alternative way.  For example, it is
quite plausible that rankers may find it easier to identify their most
and least preferred entities first rather than those entities they
place in the middle positions of their ranking
\citep{mollica2018algorithms}.  In such a scenario rankers might form
their rank ordering by first assigning their most and then least
preferred entities to a rank before filling out the middle positions
through a process of elimination using the remaining (unallocated)
entities, that is, they use a different \textit{ranking process}.  The
Extended Plackett-Luce  model relaxes the assumption of a fixed
and known ranking process.

It is somewhat natural to recast the underlying
ranking process in terms of a ``\textit{choice order}'' where the
choice order is the order in which rankers assign entities to
positions/ranks.  For example, suppose a ranker must provide a
preference ordering of $K$ entities; a choice order of
$\vec{\sigma}=(1,K,2,3,\dots,K-1)$ corresponds to the ranking process
where the ranker first assigns their most preferred entity, then their
least preferred entity before then assigning the remaining entities in
rank order from second down.  Note that the choice order $\vec{\sigma}$
is simply a permutation of the ranks~$1$ to~$K$.

Whilst the EPL model is motivated in terms of a \textit{choice
  order} as described above, we find this justification is not
always appropriate. For example, the notion of a choice order clearly
does not apply in the analysis of the Formula 1 data in
Section~\ref{sec:real_data_analyses}, where the data are simply the
finishing orders of the drivers in each race. We prefer to view the
EPL model as a flexible probabilistic model for rank ordered data;
ultimately all such probabilistic models induce a discrete
distribution $P_x$ over the set of all $K!$
permutations $\mathcal{S}_K$ and we wish this 
distribution to provide a flexible model for the observed
data. 

We adopt a
Bayesian approach to inference which we find particularly appealing
and natural as we focus heavily on predictive inference for observable
quantities. We also make three main contributions, as outlined below. When the number of entities is not small, choosing a
suitable prior distribution for~$\vec{\sigma}$, the permutation of the
ranks~$1$ to~$K$, is a somewhat daunting task. We therefore propose to
use the (standard) Plackett-Luce model to define the prior probability
of each permutation, although we note that our inference framework is
sufficiently general and does not rely on this choice. We also
  address the thorny issue of specifying informative prior beliefs
  about the entity parameters~$\lambdavec$ by proposing a class of
  priors that preserve the modal prior predictive rank 
  ordering under different choice orders~$\sigmavec$.  Constructing
suitable posterior sampling schemes for the Extended Plackett-Luce
model is challenging due to multi-modality of the marginal posterior
distribution for~$\sigmavec$, with local modes separated by large
distances within permutation space. To the best of our knowledge, the
only current solution is given by \cite{mollica2018algorithms} but
this relies on a restricted parameter space for~$\vec{\sigma}$.  In
this paper we appeal to Metropolis coupled Markov chain Monte Carlo
(MC$^3$) to overcome the difficult sampling
problem when the full parameter space for~$\vec{\sigma}$ is
considered.

The remainder of the paper is structured as follows.  In
Section~\ref{sec:model_rank_process} we outline the Extended
Plackett-Luce model and our associated notation, and in
Section~\ref{sec:interp} we provide some guidance on interpreting the
model parameters.  In Section~\ref{sec:bayes} we propose our Bayesian
approach to inference. In particular we discuss suitable choices for
the prior distribution and describe our simulation based scheme for
posterior sampling.  A simulation study illustrating the efficacy of
the posterior sampling scheme and the performance of the EPL model
over a range of values for the number of entities and number of
observations is considered in Section~\ref{sec:sim_studies}; with
further details also given in Section~\ref{sec:sim_studies_supp} of the supplementary
materials.  Section~\ref{sec:post_pred} outlines how we use the
posterior predictive distribution for inference on observable
quantities and for assessing the appropriateness of the model.  Two
real data analyses are considered in
Section~\ref{sec:real_data_analyses} to illustrate the use of the
(unrestricted) EPL model.  Section~\ref{sec:conc} offers some
conclusions.

\section{The Extended Plackett-Luce model}
\label{sec:model_rank_process}

We now present the Extended Plackett-Luce model along with our
associated notation and also discuss the interpretation of the model
parameters in terms of the preferences of entities.

\subsection{Model and notation}
\label{sec:mod}

Recall that there are~$K$ entities to be ranked and that
the collection of all entities is denoted $\mathcal{K} = \{1,\dots,K\}$.
The Extended Plackett-Luce model is only well defined for
\textit{complete} rank orderings in which all entities are included.
Thus a typical observation is $\vec{x}_i = (x_{i1},\dots,x_{iK})$
where~$x_{ij}$ denotes the entity ranked in position
$j$ in the $i$th rank ordering. 

The choice order is represented by
$\sigmavec=(\sigma_1,\ldots,\sigma_K)$, where~$\sigma_j$ denotes the
rank allocated at the $j$th stage.  Conditional on
$\sigmavec$, each entity has a corresponding parameter $\lambda_k > 0$
for $k = 1,\dots,K$; let $\vec{\lambda}=(\lambda_1,\ldots,\lambda_K)'$.
Crucially, the meaning and interpretation of $\vec{\lambda}$ depends
on $\sigmavec$ and this is addressed shortly.  

The probability of a particular rank ordering under
the Extended Plackett-Luce model \citep{MollicaT14} is defined as 
\begin{equation}
\Pr(\vec{X}_i = \vec{x}_i |\lambdavec, \sigmavec) = \prod_{j=1}^{K} \frac{\lambda_{{x_{i\sigma_{j}}}}}{\sum_{m=j}^{K} \lambda_{{x_{i\sigma_{m}}}}}.
\label{eqn:epl_prob}
\end{equation}
Therefore, the Extended Plackett-Luce probability~\eqref{eqn:epl_prob} is simply the standard
Plackett-Luce probability~\eqref{eqn:spl_prob} evaluated at ``permuted data'' $\xvec_{i}^*$
where $x_{ij}^* = x_{i\sigma_j}$ for $j=1,\dots,K$ with entity
parameters~$\lambdavec$.  Here $x_{ij}^*$ denotes the entity chosen at
the $j$th stage of the $i$th ranking process and therefore receiving
rank $\sigma_j$.

Indeed, both the (forward ranking) standard
Plackett-Luce model and (backward ranking) reverse Plackett-Luce
model  are special cases of \eqref{eqn:epl_prob} and are
recovered when~$\sigmavec=(1,\dots,K)\equiv~\mathcal{I}$, the identity permutation,  and
$\sigmavec = (K,K-1,\dots,1)$, the reverse of the identity permutation, respectively.
We use the notation $\vec{X}_i|\lambdavec, \sigmavec \sim
\text{EPL}(\lambdavec, \sigmavec)$ to denote that the probability of
rank ordering~$i$ is given by \eqref{eqn:epl_prob}.  Note that here and throughout we have
adopted different notation from that in \cite{MollicaT14} and
\cite{mollica2018algorithms} but the essential components of the model
remain unchanged.  

It is clear that the EPL probability \eqref{eqn:epl_prob} is invariant to scalar multiplication of the entity
parameters~$\lambdavec$.  This identifiability issue is not of great
concern as the parameters can be normalised as required.  However, the
parameter identifiability issue can lead to potential mixing problems
for MCMC algorithms and this is revisited in
Section~\ref{sec:postsamp}.

\subsection{Interpretation of the entity parameters $\lambdavec$}
\label{sec:interp}

A key aspect of analysing rank ordered data using Plackett-Luce type
models is the interpretation of the entity parameters~$\lambdavec$.
Moreover, it is essential to understand the interpretation of
the~$\lambda$ parameters if one is to specify informative prior beliefs about the
likely preferences of the entities.

For the Extended Plackett-Luce model,~$\lambda_k$ is proportional to
the probability that entity~$k$ is selected at the first
\textit{stage} of the ranking process and therefore ranked in
position~$\sigma_1$ of the rank ordering~$\xvec$.
Then, conditional on an entity being assigned to position~$\sigma_1$
in the rank ordering, the entity with the largest parameter of
those remaining is that most likely to be assigned to
position~$\sigma_2$, and so on.  For the
standard Plackett-Luce model, arising from the forward ranking process
with $\sigmavec=(1,\dots,K)$, we have that~$\lambda_k$ is proportional
to the probability that entity~$k$ is assigned rank~$\sigma_1=1$ (and
is thus the most preferred entity), and so on. Therefore, for the
standard Plackett-Luce model, entities with larger values are more
likely to be given a higher rank. In other words, the~$\lambda$
parameters for the standard Plackett-Luce model correspond directly
with \textit{preferences} for entities.  A consequence is that
ordering the entities in terms of their values in~$\lambdavec$, from
largest to smallest, will give the modal ordering~$\hat{\vec{x}}$,
that is, the permutation of the entities which yields the maximum
Plackett-Luce probability~\eqref{eqn:spl_prob}, given~$\lambdavec$.
Specifically, $\hat{\vec{x}}=\text{order}_{\downarrow}(\lambdavec)$,
where $\text{order}_{\downarrow}(\cdot)$ denotes the ordering
operation from largest to smallest. This makes specifying a prior
distribution for~$\lambdavec$, when $\sigmavec=(1,\dots,K)$,
relatively straightforward based on entity preferences. The
interpretation of the~$\lambda$ parameters directly in terms of
preferences can also be achieved in a straightforward manner with the
backward ranking process $\left(\sigmavec=(K,\dots,1)\right)$ of the
reverse Plackett-Luce model.  Apart from these special cases, however,
the interpretation of the~$\lambda$ parameters in terms of preferences
is not at all transparent for other choices of~$\sigmavec$.  For
example, suppose that $\lambda_i > \lambda_j$ and $\sigmavec=(2,3,1)$.
Here entity~$i$ is more likely to be ranked in second position than
entity~$j$.  Further, if another entity $\ell \neq i,j$, is assigned
to rank~$2$ then entity~$i$ is preferred for rank~$3$ ($\sigma_2$)
over entity~$j$.

Understanding the preference of the entities under the Extended
Plackett-Luce model based on values of~$\lambdavec$ and~$\sigmavec$
can be made more straightforward if we first introduce the inverse of
the choice order permutation~$\sigmavec^{-1}$. This is defined such
that $\sigmavec \circ \sigmavec^{-1} = \sigmavec^{-1} \circ \sigmavec
= \mathcal{I}$, the identity permutation, where~$\circ$ denotes composition of permutations which, in
terms of vectors, implies that if $\vec{z}=\vec{x}\circ \vec{y}$ then
$z_i=x_{y_i}$.  Here the $j$th element of $\sigmavec^{-1}$ denotes the
stage of the ranking process at which rank~$j$ is assigned.  We can
then obtain directly the modal ordering of the entities under the
EPL$(\lambdavec,\sigmavec)$ model,
$\hat{\vec{x}}^{(\sigmavec,\lambdavec)}$, and thus obtain a
representation of the preference of the entities. Here
$\hat{\vec{x}}^{(\sigmavec,\lambdavec)}$ is obtained without
enumerating any probabilities by permuting the entries in
$\hat{\vec{x}}$ (the modal ordering under the standard Plackett-Luce
model conditional on~$\lambdavec$), by~$\sigmavec^{-1}$, that is
$\hat{\vec{x}}^{(\sigmavec,\lambdavec)}=\hat{\vec{x}}\circ
\sigmavec^{-1}$. In other words, if
$\hat{\vec{x}}=\text{order}_{\downarrow}(\lambdavec)$ then
$\hat{x}^{(\sigmavec,\lambdavec)}_j=\hat{x}_{\sigma^{-1}_j}$, where~$\sigma^{-1}_j$ denotes the $j$th element of~$\sigmavec^{-1}$. Let
$\hat{\vec{x}}^{-1}$ represent the ranks assigned to the entities
under the standard Plackett-Luce model; this is obtained as the
inverse permutation corresponding to~$\hat{\vec{x}}$, that is, the
permutation such that $\hat{\vec{x}}^{-1} \circ \hat{\vec{x}} =
\mathcal{I}$, the identity permutation.  Now define
$\vec{\eta}^{(\sigmavec)}=\hat{\vec{x}}^{(\sigmavec,\lambdavec)} \circ
\hat{\vec{x}}^{-1}$; this represents the permutation of the entities
ranked under the EPL model at the stage corresponding to the rank
assigned to entities 1 to $K$ under the standard Plackett-Luce model.
It follows that, if~$\lambdavec^{(\sigmavec)}$ has $j$th element
$\lambda^{(\sigmavec)}_j=\lambdavec_{\eta^{(\sigmavec)}_j}$, where
$\eta^{(\sigmavec)}_j$ is the $j$th element of
$\vec{\eta}^{(\sigmavec)}$, then
$\hat{\vec{x}}^{(\sigmavec,\lambdavec^{\sigmavec})} \equiv
\hat{\vec{x}}$ for all $\sigmavec\in\mathcal{S}_K$, and the modal
preference ordering is preserved.

Some simplification is possible if we first order the entities in
terms of preferences. 
Clearly, if $\hat{\vec{x}}=\mathcal{I}$, the
identity permutation, then
$\hat{\xvec}^{(\sigmavec,\lambdavec)}=\sigmavec^{-1}$, and so the modal
ordering is given by the inverse choice order permutation.
Moreover, if $\hat{\vec{x}}=\mathcal{I}$ then
$\hat{\vec{x}}^{-1}=\mathcal{I}$ and so
$\vec{\eta}^{(\sigmavec)}=\sigmavec^{-1}$. It follows that choosing
$\lambdavec^{(\sigmavec)}$ such that its $j$th element is 
$\lambda^{(\sigmavec)}_j=\lambdavec_{\sigmavec^{-1}_j}$,
then $\hat{\vec{x}}^{(\sigmavec,\lambdavec^{\sigmavec})} \equiv
\mathcal{I}$ for all $\sigmavec\in\mathcal{S}_K$. Therefore if
the entities are labelled in preference order then permuting the~$\lambda$ parameters from the standard Plackett-Luce model by the
inverse of the choice order permutation will preserve the modal
permutation to be in the same preference order. This suggests a simple
strategy for specifying prior distributions for the entity parameters
which preserves modal preferences under different choice orders; we
revisit this in Section~\ref{sec:prior-lambda}.

\section{Bayesian modelling}
\label{sec:bayes}

Suppose we have data consisting of $n$ independent rank orderings, denoted
$\mathcal{D}=\{\vec{x}_1,\vec{x}_2,\ldots,\vec{x}_n\}$.  The
likelihood of $\lambdavec, \sigmavec$ is
\begin{align}
\pi(\D| \lambdavec, \sigmavec) &= \prod_{i=1}^n \Pr(\xvec_i|\lambdavec,\sigmavec) \notag \\
&= \prod_{i=1}^n \prod_{j=1}^{K} \frac{\lambda_{{x_{i\sigma_{j}}}}}{\sum_{m=j}^{K} \lambda_{{x_{i\sigma_{m}}}}}.
\label{eqn:epl_likelihood}
\end{align}
We wish to make inferences about the unknown quantities in the model
$\vec{\sigma},\lambdavec$ 
as well as future observable rank orderings $\vec{x}$. Specifically
we adopt a
Bayesian approach to inference in which we quantify our uncertainty
about the unknown quantities (before observing the data) through a
suitable prior distribution.  

\subsection{Prior specification}
\label{sec:prior}

We adopt a joint prior distribution for~$\sigmavec$ and~$\lambdavec$
of the form $\pi(\sigmavec,\lambdavec)=\pi(\lambdavec | \sigmavec)\pi(\sigmavec)$
 which explicitly emphasizes the dependence of~$\lambdavec$ on~$\sigmavec$.

\subsubsection{Prior for $\sigmavec$}
\label{sec:prior-sigma} 

For the choice ordering~$\vec{\sigma}$ we need to define a discrete
distribution~$P_{\sigmavec}$ over the~$K!$ elements
of~$\mathcal{S}_K$.  If~$K$ is not small, perhaps larger than~4,
then this could be a rather daunting task.  Given the choice order
parameter~$\sigmavec$ is a permutation, or equivalently a complete
rank ordering, one flexible option is to use the Plackett-Luce model to
define the prior probabilities for each choice order parameter.  More
specifically we let $\sigmavec |\vec{q} \sim \text{PL}(\vec{q})$
where $\vec{q} = (q_1,\dots,q_K)' \in \mathbb{R}^K_{>0}$ are to
be chosen \textit{a priori} and
\begin{equation*}
\Pr(\sigmavec|\vec{q}) = \prod_{j=1}^{K} \frac{q_{{\sigma_{j}}}}{\sum_{m=j}^{K} q_{{\sigma_{m}}}}. 
\end{equation*}

If desired, it is straightforward to assume each choice
order is equally likely \textit{a priori} by letting $q_k = q$
for $k=1,\dots,K$.  Furthermore, the inference framework that
follows is sufficiently general and does not rely on this prior
choice.  In particular, if we only wish to consider a subset of
all the possible choice orderings~$\mathcal{R}$, for example the restricted space as
in \cite{MollicaT14}, then this can be achieved by making an
appropriate choice of prior probabilities for all $\sigmavec \in
\mathcal{R}$ and letting $\Pr(\sigmavec) = 0$ for all $\sigmavec \in
\mathcal{S}_K \setminus \mathcal{R}$.  Alternatively, the Plackett-Luce prior is
sufficiently flexible that it can mimic the main features of the
restricted \cite{mollica2018algorithms} prior by suitable choice of
$\vecq$ with  
$q_{i}=q_{K+1-i}$ and $q_i > q_{i+1}$, for $i < \lceil K/2\rceil$. 

\subsubsection{Prior for $\lambdavec|\sigmavec$}
\label{sec:prior-lambda} 

It is natural to wish to specify prior beliefs in terms of preferences
for the entities. However, we have seen in Section~\ref{sec:interp}
that the interpretation of the entity parameters~$\lambdavec$ in terms
of preferences is dependent on the value of~$\sigmavec$. It follows
that specifying an informative prior for the entity parameters is
problematic unless the choice order~$\sigmavec$ is assumed to be
known.  We therefore consider separate prior distributions for~$\lambdavec$ conditional on the value of~$\sigmavec$. Since the entity
parameters $\lambda_k > 0$ must be strictly positive, a suitable,
relatively tractable, choice of conditional prior distribution is a
gamma distribution with mean $a^{(\sigmavec)}_k/b^{(\sigmavec)}_k$, that
is $\lambda_k|\sigmavec \indep \Ga(a^{(\sigmavec)}_k,b^{(\sigmavec)}_k)$
for $k=1,\dots,K$ and $\sigmavec \in \mathcal{S}_K$. Without loss of
generality we set $b^{(\sigmavec)}_k = b =1$, for all~$k$ and~$\sigmavec$ since~$b$ is not likelihood identifiable.   Our proposed strategy for
specifying the hyper-parameters~$\vec{a}^{(\sigmavec)}$ is to first
consider the prior distribution for~$\lambdavec$ under the standard
Plackett-Luce model with $\sigmavec=\mathcal{I}$. If we specify
$\vec{a}=(a_1,\ldots,a_K)'$ then
$\hat{\vec{x}}$, the modal preference ordering from the prior
predictive distribution, is
$\hat{\vec{x}}=\text{order}_{\downarrow}(\vec{a})$. Then in order to preserve the beliefs about the modal
preference ordering over different values of~$\sigmavec$ we can use
the arguments of Section~\ref{sec:interp} to specify
$a^{(\sigmavec)}_k=a_{\eta^{(\sigmavec)}_k}$ for $k=1,\ldots,K$, where
$\vec{\eta}^{(\sigmavec)}$ is as defined in Section~\ref{sec:interp}
with~$\hat{\vec{x}}$ now representing the modal preference ordering
under the prior predictive distribution conditional on
$\sigmavec=\mathcal{I}$ (the standard Plackett-Luce model). The modal
entity preferences will therefore be preserved under each value of
$\sigmavec \in \mathcal{S}_K$.  As in Section~\ref{sec:interp}, some simplification of
notation is achievable if we first re-order the entities so that
$\hat{\vec{x}}=\mathcal{I}$, in which case
$a^{(\sigmavec)}_k=a_{\sigmavec^{-1}_k}$ for $k=1,\ldots,K$.  Clearly,
letting $a_k=a$ for all~$k$ induces a
uniform prior predictive distribution over all preference orders
(irrespective of the choice order~$\sigmavec$). Such a prior
represents the situation where we are unwilling to  favour any particular
preference ordering \textit{a priori}.

\subsection{Bayesian model}
\label{sec:fullbayesmodel}

The complete Bayesian model is
\begin{align*}
\hspace{3cm} \vec{X}_i|\lambdavec, \sigmavec &\indep \text{EPL}(\lambdavec, \sigmavec),
 &i=1,\ldots,n,\\
\lambda_k|\sigmavec,\vec{a} &\indep \Ga(a^{(\sigmavec)}_k,1), &k=1,\dots,K,\\
\sigmavec|\vecq &\mysim \text{PL}(\vecq),
\end{align*}
that is, we assume that our observations follow the distribution
specified by the Extended Plackett-Luce model~\eqref{eqn:epl_prob} and
the prior distribution for $(\lambdavec,\sigmavec)$ is as described in
Section~\ref{sec:prior}.

The full joint density of all stochastic quantities in the model (with dependence
on fixed hyper-parameters suppressed) is
\[
\pi(\vec{\sigma},\vec{\lambda},\mathcal{D})=\pi(\mathcal{D}|\vec{\lambda},\vec{\sigma})\pi(\vec{\lambda}|
\vec{\sigma})\pi(\vec{\sigma}). 
\]
From which we quantify our beliefs about $\sigmavec$ and $\lambdavec$
through their joint posterior density
\[
\pi(\vec{\sigma},\vec{\lambda}|\mathcal{D}) \propto \pi(\mathcal{D}|\vec{\lambda},\vec{\sigma})\pi(\vec{\lambda}|
\vec{\sigma})\pi(\vec{\sigma})
\]
which is obtained via Bayes' Theorem. The posterior density
$\pi(\vec{\sigma},\vec{\lambda}|\mathcal{D})$
is not available in closed form and so we use simulation-based
methods to sample from the posterior distribution  as described in the next section.

\subsection{Posterior sampling}
\label{sec:postsamp}

Due to the complex nature of the posterior distribution we use Markov
chain Monte Carlo (MCMC) methods in order to sample realisations from
$\pi(\vec{\sigma},\vec{\lambda}|\mathcal{D})$.
The structure of the model lends itself naturally to consider sampling
alternately from two blocks of full conditional distributions:
$\pi(\sigmavec|\lambdavec,\mathcal{D})$ and $\pi(\lambdavec|\sigmavec,\mathcal{D})$.

\subsubsection{Sampling the choice order parameter $\sigmavec$ from $\pi(\sigmavec|\lambdavec,\mathcal{D})$}
\label{sec:choice_order_inf}

Given the choice order parameter~$\sigmavec$ is a member of~$\mathcal{S}_K$ it is fairly straightforward to obtain its (discrete) full conditional distribution; specifically this is the discrete distribution with probabilities 
\begin{align*}
\Pr(\sigmavec = \sigmavec_j|\lambdavec,\D) \propto \pi(\D|\lambdavec, \sigmavec = \sigmavec_j) \pi(\vec{\lambda}|
\vec{\sigma} = \vec{\sigma}_j)\Pr(\sigmavec = \sigmavec_j)
\end{align*}
for $j=1,\dots,K!$.
Clearly sampling from this full conditional will require~$K!$
evaluations of the EPL likelihood $\pi(\D|\lambdavec, \sigmavec =
\sigmavec_j)$ and so sampling from $\Pr(\sigmavec =
\sigmavec_j|\lambdavec,\D)$ for $j=1,\dots,K!$ (a Gibbs update) is probably only plausible if~$K$ is sufficiently small; perhaps not much greater than~5.
Of course, the probabilities $\Pr(\sigmavec = \sigmavec_i|\lambdavec,\D)$ and $\Pr(\sigmavec = \sigmavec_j|\lambdavec,\D)$ are conditionally independent for $i\neq j$ and so could be computed in parallel which may facilitate this approach for slightly larger values of~$K$.

So as to free ourselves from the restriction to the case where~$K$ is small we instead consider a more general sampling strategy by constructing a Metropolis-Hastings proposal mechanism for updating~$\sigmavec$.
Our investigation into the likelihood of the Extended Plackett-Luce
model given different choice orders in Section~\ref{sec:likeli_info_choice_order} of the supplementary
material revealed that $\pi(\D|\lambdavec,\sigmavec)$ is likely to be multi-modal.
Further, local modes can be separated by large distances within permutation space.
In an attempt to effectively explore this large discrete space we consider~$5$ alternative proposal mechanisms; each of which occurs with probability $p_\ell$ for $\ell=1,\dots,5$.
The~$5$ mechanisms to construct the proposed permutation~$\sigmavec^\dagger$ are as follows.

\begin{enumerate}
\item The random swap: sample two positions $\phi_1,\phi_2 \in \{1,\dots,K \}$ uniformly at random and let the proposed choice order~$\sigmavec^\dagger$ be the current choice order~$\sigmavec$ where the elements in positions~$\phi_1$ and~$\phi_2$ have been swapped.
\item The Poisson swap: sample $\phi_1 \in \{1,\dots,K \}$ uniformly at random and let~$\phi_2=\phi_1+m$ where $m=(-1)^\tau f$, $\tau \sim \Bern(0.5)$ and $f \sim \Po(t)$. Note that~$t$ is a tuning parameter and $\phi_2 \to \{(\phi_2-1) \bmod K\} + 1$ as appropriate. Again the proposed choice order~$\sigmavec^\dagger$ is formed by swapping the elements in positions~$\phi_1$ and~$\phi_2$ of the current choice order~$\sigmavec$.
\item The random insertion \citep{bezakova2006graph}: sample two positions $\phi_1 \neq \phi_2 \in \{1,\dots,K \}$ uniformly at random and let the proposed choice order~$\sigmavec^\dagger$ be formed by taking the value in position~$\phi_1$ and inserting it back into the permutation so that it is instead in position~$\phi_2$.
\item The prior proposal: here~$\sigmavec^\dagger$ is simply an independent draw from the prior distribution, that is, $\sigmavec^\dagger|\vecq \sim \text{PL}(\vecq)$.
\item The reverse proposal: here~$\sigmavec^\dagger$ is defined to be the reverse ordering of the current permutation~$\sigmavec$, that is, $\sigmavec^\dagger = \sigmavec_{K:1} = (\sigma_K,\dots,\sigma_1).$
\end{enumerate}

Note that performing either of the swap or insertion moves (1--3) above may result in slow exploration of the set of all permutations as the proposal~$\sigmavec^\dagger$ may not differ much from the current value~$\sigmavec$.
To alleviate this potential issue we propose to iteratively perform each of these moves~$S$ times, where~$S$ is to be chosen (and fixed) by the analyst.
More formally (when using proposal mechanisms 1--3) we construct intermediate proposals~$\sigmavec^\dagger_s$ from the ``current'' choice order~$\sigmavec^\dagger_{s-1}$ for $s=1,\dots,S$.
Here~$\sigmavec^\dagger_{0} = \sigmavec$ and the proposed value for which we evaluate the acceptance probability is $\sigmavec^\dagger = \sigmavec^\dagger_S$.
Further, for moves~1 and~2 it may seem inefficient to allow for the ``null swap'' $\phi_1=\phi_2$, however this is done to avoid only proposing permutations with the same (or opposing) parity as the current value.
Put another way, as $S \to \infty$ we would expect $\Pr(\sigmavec^\dagger|\sigmavec) > 0$ for all $\sigmavec^\dagger,\sigmavec \in \mathcal{S}_K$ and this only holds if we allow for the possibility that $\phi_1=\phi_2$.
Finally we note that each of these proposal mechanisms is ``simple'' in the respect that $\Pr(\sigmavec^\dagger|\sigmavec) = \Pr(\sigmavec|\sigmavec^\dagger)$ and so the proposal ratio cancels in each case.
The full acceptance ratio is presented within the algorithm outline in Section~\ref{sec:epl_mc3_alg}.

\subsubsection{Sampling the entity parameters $\lambdavec$ from
  $\pi(\lambdavec|\sigmavec,\mathcal{D})$}
\label{sec:lambda-samp}

Bayesian inference for variants of standard Plackett-Luce models
typically proceeds by first introducing appropriate versions of the
latent variables proposed by \cite{caron2012efficient}, which in turn
facilitate a Gibbs update for each of the entity parameters (assuming
independent Gamma prior distributions are chosen).
However we found that this strategy does not work well for entity parameter inference under the Extended Plackett-Luce model (not reported here).
We therefore propose to use a Metropolis-Hastings step for sampling
the entity parameters, specifically we use (independent) log normal
random walks for each entity parameter in turn and so the proposed
value is $\lambda^\dagger_k \indep \LN(\log \lambda_{k},
\sigma_{k}^2)$ for $k=1,\dots,K$. We also implement a rescaling
  step in the MCMC scheme, analogous to that in \cite{caron2012efficient}, in order to mitigate
  the poor mixing that is caused by the invariance of the Extended Plackett-Luce likelihood to scalar multiplication of the $\lambdavec$ parameters.
Full details are given in
Section~\ref{sec:epl_mc3_alg}.

\subsubsection{Metropolis coupled Markov
chain Monte Carlo}
\label{sec:pt}

Unfortunately the sampling strategy described above in
Sections~\ref{sec:choice_order_inf} and \ref{sec:lambda-samp}   proves
ineffective when~$K$ is not small, with the Markov chain suffering
from poor mixing, 
particularly for~$\sigmavec$  where the chain is prone
to becoming stuck in local modes (results not reported here).  In an
attempt to resolve these issues, and therefore aid the exploration of
the posterior distribution we appeal to Metropolis coupled Markov
chain Monte Carlo, or \textit{parallel tempering}. 

Metropolis coupled Markov chain Monte Carlo \citep{geyer1991markov}, is a sampling technique that aims to improve the mixing of Markov chains in comparison to standard MCMC methods particularly when the target distribution is multi-modal \citep{gilks1996strategies,brooks1998markov}.
The basic premise is to consider~$C$ chains evolving simultaneously, each of which targets a tempered posterior distribution $\tilde\pi_c(\theta_c|\D)\propto \pi(x|\theta_c)^{1/T_c} \pi(\theta_c)$, where~$T_c \geq 1$ is the temperature of chain~$c$, and $\theta_c=\{\sigmavec_c,\lambdavec_c\}$.
Note that the posterior of interest is recovered when $T_c=1$.
Further note that we have only considered a tempered likelihood component as we suggest that any prior beliefs should be consistent irrespective of the model choice.
Now, as the posteriors $\tilde\pi_c(\theta_c|\D)$ are conditionally independent given~$\D$, we can consider them to be targeting the joint posterior
\begin{equation}
\pi(\theta_1,\dots,\theta_C|\D) = \prod_{c=1}^C \tilde\pi_c(\theta_c|\D).
\label{eqn:mc3_joint_target}
\end{equation}
Suppose now we propose to swap~$\theta_i$ and~$\theta_j$ for some $i\neq j$ within a Markov chain targeting the joint posterior~(\ref{eqn:mc3_joint_target}).
If we let $\thetavec=(\theta_1,\dots,\theta_C)$ denote the current state and $\thetavec^\dagger=(\theta_1^\dagger,\dots,\theta_C^\dagger)$ the proposed state where $\theta_i^\dagger=\theta_j$, $\theta_j^\dagger=\theta_i$ and $\theta_\ell^\dagger=\theta_\ell$ for $\ell \neq i,j$.
Then, assuming a symmetric proposal mechanism, the acceptance probability of the state space swap is $\min(1,A)$ where
\begin{equation*}
A = \frac{ \pi(\D|\theta_j)^{1/T_i} \pi(\D|\theta_i)^{1/T_j} }{\pi(\D|\theta_i)^{1/T_i} \pi(\D|\theta_j)^{1/T_j} }.
\end{equation*}
Of course, if the proposal mechanism is not symmetric then the probability~$A$ must be multiplied by the proposal ratio $q(\thetavec|\thetavec^\dagger)/q(\thetavec^\dagger|\thetavec)$.
Further, it is straightforward to generalise the above acceptance probability to allow the states of more than~$2$ chains to be swapped.
However, this is typically avoided as such a proposal can
result in poor acceptance rates. Our specific Metropolis coupled Markov chain Monte Carlo algorithm is
outlined 
in the next section. 

\subsection{Outline of the posterior sampling algorithm}
\label{sec:epl_mc3_alg}
A parallel Metropolis coupled Markov chain Monte Carlo algorithm to
sample from the joint posterior distribution of the skill parameters~$\lambdavec$ and the choice order parameter~$\sigmavec$ is as follows.

\begin{enumerate}
\item Tune:
\begin{itemize}
\item choose the number of chains ($C$); let $T_1 =1$ and choose $T_c > 1$ for $c=2,\dots,C$ 
\item choose appropriate values for the MH proposals outlined in Sections~\ref{sec:choice_order_inf} and \ref{sec:lambda-samp}
\end{itemize}  
\item Initialise: take a prior draw or alternatively choose $\sigmavec_c \in \mathcal{S}_K$ and $\lambdavec_c \in \mathbb{R}^K_{>0}$ for $c=1,\dots,C$
\item For $c=1,\dots,C$ perform (in parallel) the following steps:
\begin{itemize}
\item For $k=1,\dots,K$
\begin{itemize}
\item draw $\lambda_{ck}^\dagger|\lambda_{ck} \sim \LN(\log \lambda_{ck}, \sigma_{\lambda_{ck}}^2)$
\item let $\lambda_{ck} \to \lambda_{ck}^\dagger$ with probability $\min(1,A)$ where
\[
A = \left\{ \frac{\pi(\D|\lambdavec_{c,-k}, \lambda_{ck} = \lambda_{ck}^\dagger,\sigmavec_c)} 
{\pi(\D|\lambdavec_c,\sigmavec_c)} \right\}^{1/T_c} \times \left( \frac{\lambda_{ck}^\dagger}{\lambda_{ck}}\right)^{a^{(\sigmavec)}_k} e^{(\lambda_{ck} - \lambda_{ck}^\dagger)}
\]
\end{itemize}
\item Sample $\ell$ from the discrete distribution with probabilities $\Pr(\ell=i)=p_{i,c}$ for $i=1,\dots,5$
\begin{itemize}
\item propose $\sigmavec^\dagger_c$ using proposal mechanism~$\ell$
\item let $\sigmavec_c \to \sigmavec^\dagger_c$ with probability $\min(1,A)$ where 
\[
A = \left\{ \bigfrac{\pi(\D|\lambdavec_c,\sigmavec^\dagger_c)}{\pi(\D|\lambdavec_c,\sigmavec_c)} \right\}^{1/T_c} \times
\bigfrac{\pi(\lambdavec_c|\sigmavec^\dagger_c)}{\pi(\lambdavec_c|\sigmavec_c)}
 \bigfrac{\Pr(\sigmavec^\dagger_c)}{\Pr( \sigmavec_c)}
\]
\end{itemize}
\vspace{0.5cm}

\item Rescale
\begin{itemize}
\item sample $\Lambda_c^\ddagger \sim \Ga \left(\sum\limits_{k=1}^K a^{(\sigmavec)}_k, 1 \right)$.
\item calculate $\Sigma_c = \sum\limits_{k=1}^{K}  \lambda_{ck}$.
\item let $\lambda_{ck} \rightarrow  \lambda_{ck}\;  \Lambda_c^\ddagger / \Sigma_c$ for $k=1,\dots,K$.
\end{itemize}
\end{itemize}
\item Sample a pair of chain labels ($i,j$) where $1 \leq i\neq j \leq C$
\begin{itemize}
\item let $(\lambdavec_i,\sigmavec_i) \to (\lambdavec_j,\sigmavec_j)$ and $(\lambdavec_j,\sigmavec_j) \to (\lambdavec_i,\sigmavec_i)$  with probability $\min(1,A)$ where
\[
\hspace{-2.5cm} A = \frac{ \pi(\D|\lambdavec_j,\sigmavec_j)^{1/T_i} \pi(\D|\lambdavec_i,\sigmavec_i)^{1/T_j} }{\pi(\D|\lambdavec_i,\sigmavec_i)^{1/T_i} \pi(\D|\lambdavec_j,\sigmavec_j)^{1/T_j} }
\]
\end{itemize}
\item Return to Step~3.
\end{enumerate}

\subsubsection{Tuning the MC$^3$ algorithm}
\label{sec:tuning_mc3}

The Metropolis coupled Markov chain Monte Carlo
scheme targets the joint density~\eqref{eqn:mc3_joint_target} by simultaneously evolving~$C$ chains; each of which targets an
alternative (tempered) density~$\tilde\pi_c(\theta_c|\D)$.
Given the data~$\D$, these chains are conditionally independent and should therefore be individually tuned to target their respective density in a typical fashion.
Of course, it may not be possible to obtain near optimal acceptance rates within the posterior chain (and other chains with temperatures $\simeq 1$) however the analyst should aim to ensure reasonable acceptance rates; even if this results in small moves around the parameter space.
Tuning the between chain proposal
(Step~4 of the MC$^3$ algorithm) can be tricky in general. 
The strategy we
suggest, also advocated by \cite{wilkinsonmc3blog}, is that where the
temperatures are chosen such that they exhibit geometric spacing, that is, $T_{c+1}/T_{c} = r$ for some $r > 1$; this eliminates the burden of
specifying~$C-1$ temperatures and instead only requires a choice of~$r$. 
We also suggest only considering swaps between adjacent chains as  intuitively the target densities are most similar when $|T_c
- T_{c+1}|$ is small.
 It is generally accepted that
between chain acceptance rates of around~$20\%$ to~$60\%$ provide
reasonable mixing (with respect to the joint density of
$\theta_1,\dots,\theta_C$); see, for example,
\cite{geyer1995annealing,altekar2004parallel}.
 A
suitable choice of the temperature ratio~$r$ can be guided via pilot
runs of the MC$^3$ scheme and  individual temperatures can also be
adjusted as appropriate.

\section{Simulation study}
\label{sec:sim_studies}

To investigate the performance of the posterior sampling algorithm
outlined in Section~\ref{sec:epl_mc3_alg} we apply it on several synthetic datasets.
We consider $K \in \{ 5,10,15,20\}$ entities and generate $500$
 rank orderings for each choice of $K$.
Further we subset each of these datasets by taking the first $n \in \{
20,50,200,500\}$ orderings thus giving rise to~16 (nested) datasets.
The parameter values $(\lambdavec',\sigmavec')$ from which these data are generated are drawn from the prior distribution outlined in Section~\ref{sec:prior} with $a_k=q_k=1$ for $k=1,\dots,K$.
That is, all choice orders and entity preferences (specified by the $(\lambdavec',\sigmavec')$ pair) are equally likely.
The values of the parameters for each choice of~$K$ are given in Section~\ref{sec:sim_studies_supp} of the supplementary materials.
For each dataset, posterior samples were obtained via the algorithm outlined in Section~\ref{sec:epl_mc3_alg}.
We choose to use~$C=5$ chains in each case, with both the temperatures and tuning parameters chosen appropriately.
The raw posterior draws are also thinned to obtain (approximately)
$10$K un-autocorrelated draws from the posterior distribution.  Note that standard MCMC diagnostics
were applied to the (continuous) parameters~$\lambda$ and also the
(log) observed data likelihood~\eqref{eqn:epl_likelihood}. To
alleviate potential concerns about the sampling  of the discrete
choice order parameter~($\sigmavec$) we checked that the marginal
posterior distribution $\pi(\sigmavec|\D)$ was consistent under
multiple runs of our algorithm.

\begin{figure}[t]
\begin{center}
\begin{minipage}[b]{0.32\linewidth}
        \centering $n=20,K=15$
 	    \includegraphics[width=.99\linewidth, clip, trim= 0 17 20 55]{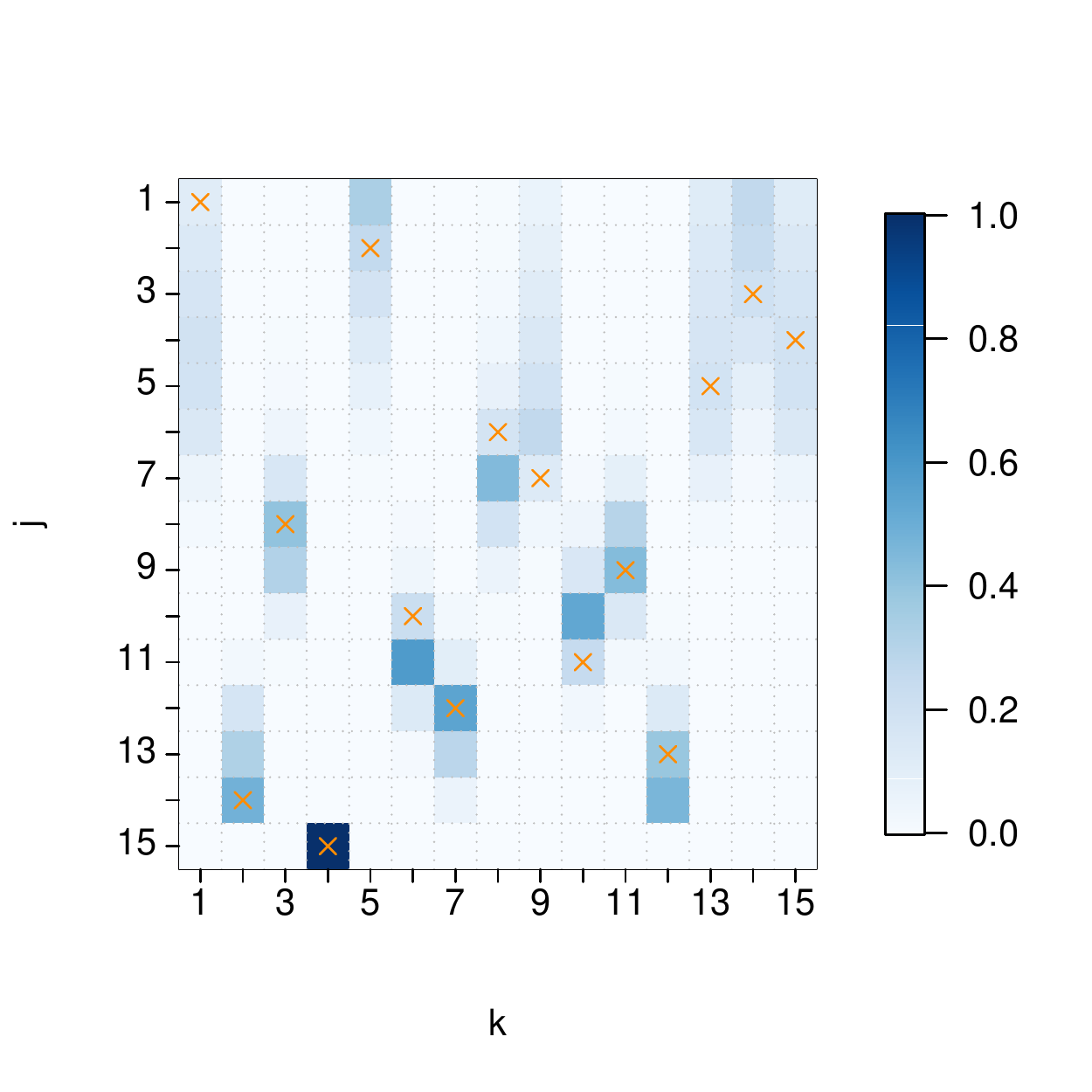}
\end{minipage} %
\begin{minipage}[b]{0.32\linewidth}
        \centering  $n=20,K=20$
        \includegraphics[width=.99\linewidth, clip, trim=0 17 20 55]{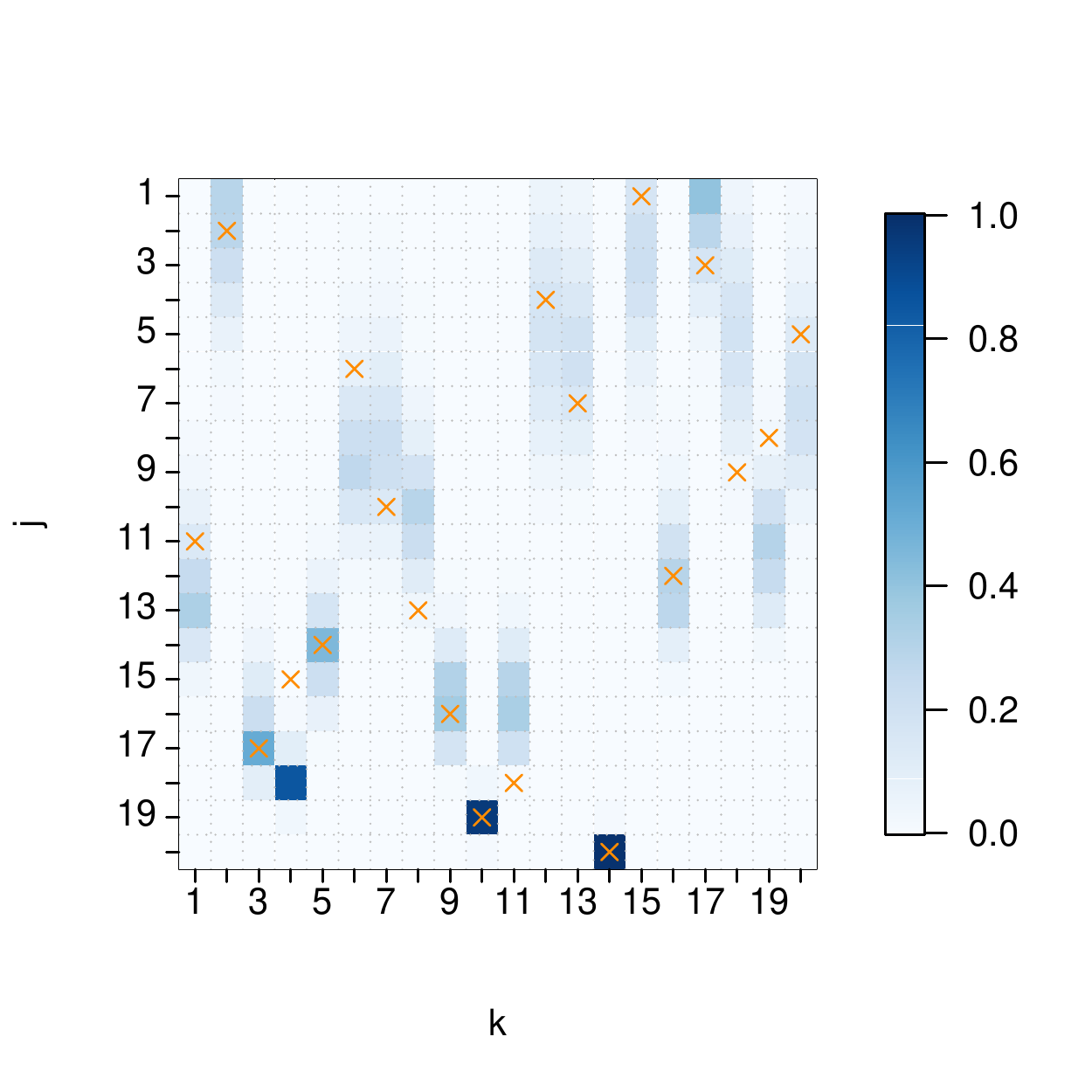}
\end{minipage}
\begin{minipage}[b]{0.32\linewidth}
        \centering  $n=50,K=20$
        \includegraphics[width=.99\linewidth, clip, trim=0 17 20 55]{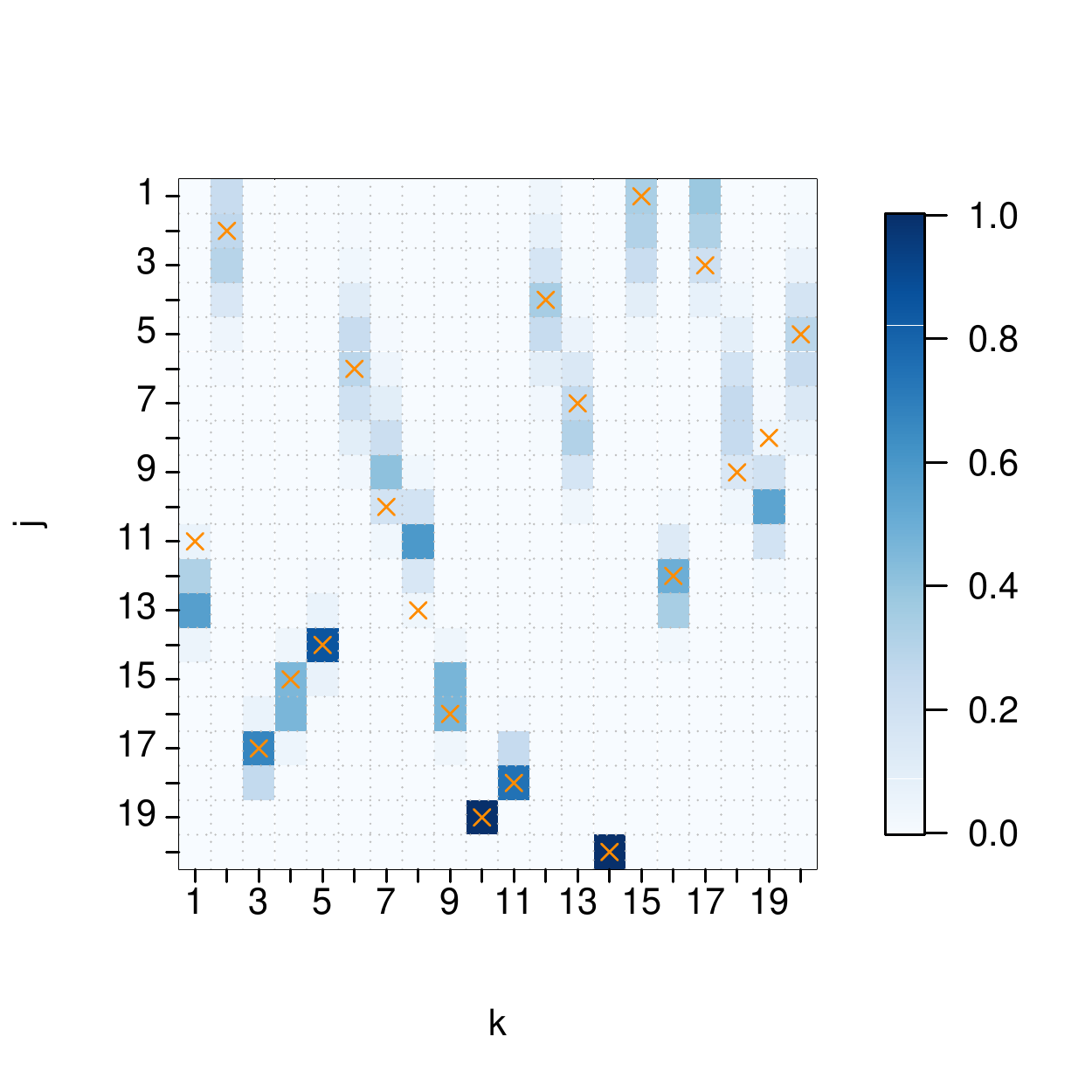}
\end{minipage}
\caption{Synthetic data: heat maps of $\Pr(\sigma_j=k|\D)$ for those analyses where~$\sigmavec'$ was not observed; the crosses highlight $\Pr(\sigma_j=\sigma'_j|\D)$ in each case.}
\label{fig:sim_study_images_paper}
\end{center}
\end{figure}

Table~\ref{tab:sim_study} shows the posterior probability $\Pr(\sigmavec'|\D)$ of the choice order parameter used to generate each respective dataset.
\begin{table}[b!]
\spacingset{1.3}
\scriptsize
\begin{center}
\begin{tabular}{cc|*{4}c}
\multicolumn{2}{c}{ } & \multicolumn{4}{c}{$n$} \\
$K$&  & 20 & 50 & 200 & 500\\
\hline
5 & $\Pr(\sigmavec'|\D)$& $0.294^*$ & $0.716^*$ & $1.000^*$ & $1.000^*$\\
&$\frac{1}{K}\sum_k \left[ \E(\log\lambda_k|\D) - \log\lambda_k'\right]^2$ & $0.045$ & $0.028$ & $0.030$ & $0.009$ \\
\hline
10 &$\Pr(\sigmavec'|\D)$& $0.156^*$ & $0.604^*$ & $1.000^*$ & $1.000^*$\\
&$ \frac{1}{K} \sum_k \left[ \E(\log\lambda_k|\D) - \log\lambda_k'\right]^2$ & $0.276$ & $0.059$ & $0.030$ & $0.006$ \\
\hline
15 & $\Pr(\sigmavec'|\D)$& $0.000$ & $0.006$ & $0.072$ & $0.548^*$\\
&$ \frac{1}{K}\sum_k \left[ \E(\log\lambda_k|\D) - \log\lambda_k'\right]^2$ & --- & $0.020$ & $0.010$ & $0.002$ \\
\hline
20 & $\Pr(\sigmavec'|\D)$ & $0.000$ & $0.000$ & $0.035$ & $0.313^*$ \\
&$\frac{1}{K}\sum_k \left[ \E(\log\lambda_k|\D) - \log\lambda_k'\right]^2$ & --- & --- & $0.010$ & $0.009$ \\
\end{tabular} 
\caption{Posterior probability $\Pr(\sigmavec'|\D)$ of the choice order used to generate each dataset along with mean squared error between the (log) values~$\lambdavec'$ used to generate the data and the posterior expectation of the (log) entity parameters conditional on $\sigmavec'$. $^*$indicates that~$\sigmavec'$ is also the (posterior) modal observed choice order.}
\label{tab:sim_study}
\end{center}
\end{table} 
Perhaps unsurprisingly we see that for each $K \in \{ 5,10,15,20\}$
the posterior support for the choice order parameter used to generate
the data increases with the number of observations (rank orderings) considered, that is, $\Pr(\sigmavec'|\D) \to 1$ as $n \to \infty$.
Interestingly we observe reasonable posterior support for~$\sigmavec'$ when only considering $n=20$ preference orders of $K=10$ entities.
However for some of the analyses, those where~$n$ is relatively small in comparison to~$K$, the choice order~$\sigmavec'$ is not observed in any of the~$10$K posterior draws.
Further inspection of the marginal posterior (of $\sigmavec$) for these analyses reveals that there is a large amount of uncertainty on the choice order parameter.
That said, the posterior draws of~$\sigmavec$ are reasonably consistent with the~$\sigmavec'$ used to generate the respective datasets; this can be seen by considering the marginal posterior distribution for each stage in the ranking process, that is, $\Pr(\sigma_j=k|\D)$ for $j,k \in\{1,\dots,K\}$.
Figure~\ref{fig:sim_study_images_paper} shows heat maps of $\Pr(\sigma_j=k|\D)$ for those analyses where~$\sigmavec'$ was not observed; the crosses highlight $\Pr(\sigma_j=\sigma'_j|\D)$ in each case.
These figures reveal that, even with limited information, we are able
to learn the lower entries in~$\sigmavec$ fairly well and much of the
uncertainty resides within the first few stages of the ranking
  process.
Section~\ref{sec:sigma_positions_supp} of the supplementary materials presents the $\Pr(\sigma_j=k|\D)$ from Figure~\ref{fig:sim_study_images_paper} in tabular form along with the image plots for the remaining analyses.

For the Extended Plackett-Luce model we are not only trying to
quantify our uncertainty about the choice order parameter but also
about the entity parameters.
As discussed in Section~\ref{sec:model_rank_process} the entity
parameter values~$\lambdavec$ only have a meaningful interpretation for a given choice order parameter~$\sigmavec$.
That said, the values of the entity parameters are of little interest here and so we instead consider the mean squared error between the (log) values~$\lambdavec'$ used to generate the data and the posterior expectation of the (log) entity parameters (conditional on the $\sigmavec'$ used to generate the data).
Table~\ref{tab:sim_study} therefore shows $ \frac{1}{K} \sum_k \left[ \E_{\lambda_k,\sigmavec=\sigmavec'|\D}(\log\lambda_k|\D) - \log\lambda_k'\right]^2$ from which we see that, in general, the inferred posterior means agree with the values~$\lambdavec'$ used the generate the data.
Of course, there is also uncertainty on these parameters;
Section~\ref{sec:sim_studies_supp} of the supplementary materials
contains boxplots of the marginal posterior distributions of
$\log\lambda$ and these show that there is reasonable posterior
support for~$\lambdavec'$, even when~$n$ is small relative to $K$.
Naturally we can not obtain $\E_{\lambdavec,\sigmavec=\sigmavec'|\D}(\log\lambdavec)$ for those analyses where~$\sigmavec'$ is not observed.
However, although prohibitive for~$\lambdavec$ inference, this does not prohibit inferences on observable quantities (rank orders) as this is achieved via the posterior predictive distribution; this is the topic of the next section.

\section{Inference and model assessment via the posterior predictive distribution}
\label{sec:post_pred}

In this section we consider methods for performing inference for the
entities by appealing to the posterior predictive distribution which
will also provide us with a mechanism for detecting lack of model fit.
We also outline methods for obtaining the mode of the posterior
predictive distribution when the number of entities is large.  By
definition the modal ranking (from the posterior predictive
distribution) is that which is most likely given the data and so this
can be thought of as the \textit{aggregate} ranking
\citep{johnson2018revealing} from a rank
aggregation perspective if desired.

\subsection{Inference for entity preferences}
\label{sec:inf-ent}

The Extended Plackett-Luce model is only defined for complete rankings
and so the posterior predictive distribution is a discrete
distribution defined over all possible observations~$\tilde{\vec{x}} 
\in \mathcal{S}_K$.  It is straightforward to approximate these
probabilities by taking the expectation of the EPL
probability~\eqref{eqn:epl_prob} over the posterior distribution for
$\lambdavec,\sigmavec$. Specifically the posterior predictive
probability of any observation $\tilde{\vec{x}}$ is
$\Pr(\tilde{\vec{x}}|\D) \simeq \text{E}_{\lambdavec,\sigmavec|\D}
\left[\Pr(\tilde{\vec{x}} |\lambdavec, \sigmavec) \right]$ where the
approximation is exact in the limit of infinite posterior samples, and
where $\Pr(\tilde{\vec{x}} |\lambdavec, \sigmavec)$ is given in Equation~\eqref{eqn:epl_prob}.  It
follows that, in principle, we can obtain the full posterior
predictive distribution by simply computing $\Pr(\tilde{\vec{x}}|\D)$
for each of the~$K!$ possible observations $\tilde{\vec{x}}\in\mathcal{S}_K$.  We can then use this
distribution, for example, to obtain the marginal posterior predictive
probability that entity~$k$ is ranked in position $j$, that is, $\Pr(\tilde{x}_j=k|\D)$ for $j,k \in
\{1,\dots,K \}$.  Further, the modal ordering~$\hat{\vec{x}}$ is also straightforward
to obtain and is simply that which has largest posterior predictive probability.
However, when the number of entities is larger than say~$9$, this
procedure involves enumerating the predictive probabilities for more
than~$\mathcal{O}(10^6)$ possible observations.  Clearly this becomes
computationally infeasible as the number of entities increases;
particularly as computing the posterior predictive probability also involves
taking the expectation over many thousands of posterior draws.  When
the number of entities renders full enumeration infeasible we suggest
approximating the posterior predictive distribution via a Monte Carlo based
approach as in \cite{johnson2018revealing}.  In particular we obtain a
collection~$P=\left\{\tilde{\vec{x}}^{(m)}_\ell\right\}_{m=1,
  \ell=1}^{m=M, \ell=L}$
 of draws from the posterior predictive
distribution by sampling~$L$ rank orderings at each iteration of the~$M$
iterations of the posterior sampling 
scheme.  We can then approximate $\Pr(\tilde{x}_j=k|\D)$ by the
empirical probability computed from the collection of
rankings~$P$, that is
$\hat{\Pr}(\tilde{x}_j=k|\mathcal{D})=\frac{1}{ML}\sum_{m=1}^{M}\sum_{\ell=1}^{L}\mathbb{I}(\tilde{x}^{(m)}_{\ell
  j}=k)$, where $\mathbb{I}(x)$ denotes an indicator function which
returns 1 if $x$ is true and 0 otherwise.  Finally, in order to find the mode of the
posterior predictive distribution we propose using an efficient optimisation algorithm based on cyclic coordinate
ascent; full details are provided in \cite{johnson2018revealing}.

\subsection{Model assessment via posterior predictive checks}
\label{sec:gof}

In the Bayesian framework assessment of model fit to the data can be
provided by comparing observed quantities with potential future
observations through the posterior predictive distribution; the basic
idea being that the observed data~$\D$ should appear to be a plausible
realisation from the posterior predictive distribution.  This approach
to Bayesian goodness of fit dates back at least to \cite{Guttman67}
and is described in detail in \cite{GelmanBDA3}, for example.  Several
methods for assessing goodness of fit for models of rank ordered data
were proposed in \cite{CohenMallows83} and more recently similar
methods have been developed in a Bayesian framework by, amongst
others, \cite{YaoB99}, \cite{MollicaT17}, \cite{johnson2018revealing}
and, specifically for the Extended Plackett Luce model, by
\cite{mollica2018algorithms}.  In the illustrative examples on real
data in Section~\ref{sec:real_data_analyses} we propose a range of
diagnostics tailored to the specific examples.  For example, one
generic method for diagnosing lack of model fit is to monitor the
(absolute value of the) discrepancy between the marginal posterior
predictive probabilities of entities taking particular ranks with the
corresponding empirical probabilities computed from the observed data.
That is, we consider $d_{jk}=|\Pr(\tilde{x}_j=k|\D) -
\Pr(x_j=k)|$ where
$\Pr(x_j=k)=\frac{1}{n}\sum_{i=1}^{n}\mathbb{I}(x_{ij}=k)$ is
computed from those $\vec{x} \in \D$ and the posterior predictive
probabilities $\Pr(\tilde{x}_j=k|\D)$ are computed as described in
Section~\ref{sec:inf-ent}.  These discrepancies $d_{jk}$ for
$j,k \in \{1,\ldots,K\}$ can then be depicted as a heat map where large values
could indicate potential lack of model fit.  By focusing on the
marginal probabilities $\Pr(x_j=k)$ we obtain a broad-scale
``first-order'' check on the model, but, as described in
\cite{CohenMallows83}, we could also look at finer-scale features such
as pairwise comparisons, triples and so on.  Of course, if the full
posterior predictive distribution over all $K!$ possible observations
is available (that is, if $K$ is small) then we could compare the
empirical distribution with the posterior predictive distribution
directly; this is considered in the example in Section~\ref{sec:song}.

\section{Illustrative examples}
\label{sec:real_data_analyses}

We now summarise analyses of two real datasets which together highlight
how valuable insights can be obtained by considering the Extended Plackett-Luce model as opposed to simpler alternatives.
Our conclusions are compared to those obtained under a standard
Plackett-Luce analysis; here posterior samples are obtained using the
Gibbs sampling scheme of \cite{caron2012efficient}.

\subsection{Song data}
\label{sec:song}

For our first example we consider a dataset with a long standing in the literature that was first presented in \cite{critchlow1991probability}.
The original dataset was formed by asking ninety-eight students to rank $K=5$ words, (1) \emph{score}, (2) \emph{instrument}, (3) \emph{solo}, (4) \emph{benediction} and (5) \emph{suit}, according to the association with the target word ``\emph{song}''.
However, the available data given in \cite{critchlow1991probability} is in grouped format
and the ranking of 15 students are unknown and hence discarded.
The resulting dataset therefore comprises $n=83$ ranking orderings and is reproduced in the supplementary materials.

Posterior samples are obtained via the algorithm outlined in
Section~\ref{sec:epl_mc3_alg} where the prior specification is as in
Section~\ref{sec:prior} with $q_k = 1$ and $a_k=1$ (for
$k=1,\dots,K$) and so all choice and preference orderings are equally likely \textit{a priori}.
The following results are based on a typical run of our (appropriately
tuned) MC$^3$ scheme initialised from the prior, with appropriate
burn-in and thin to obtain 10K (almost) un-autocorrelated realisations
from the posterior distribution. As in the simulation studies we check that $\pi(\sigmavec|\D)$ is consistent under multiple runs of our algorithm and also use standard MCMC diagnostics on the~$\lambda$ parameters and the (log) observed data likelihood~\eqref{eqn:epl_likelihood}. 
The algorithm runs fairly quickly, with~C code on a five threads of an Intel Core i7-4790S CPU
(3.20GHz clock speed) taking around~18 seconds.

\begin{figure}[b!]
\begin{center}
\begin{minipage}[b]{0.35\linewidth}
        \centering EPL
 	    \includegraphics[width=.99\linewidth, clip, trim= 0 17 10 55]{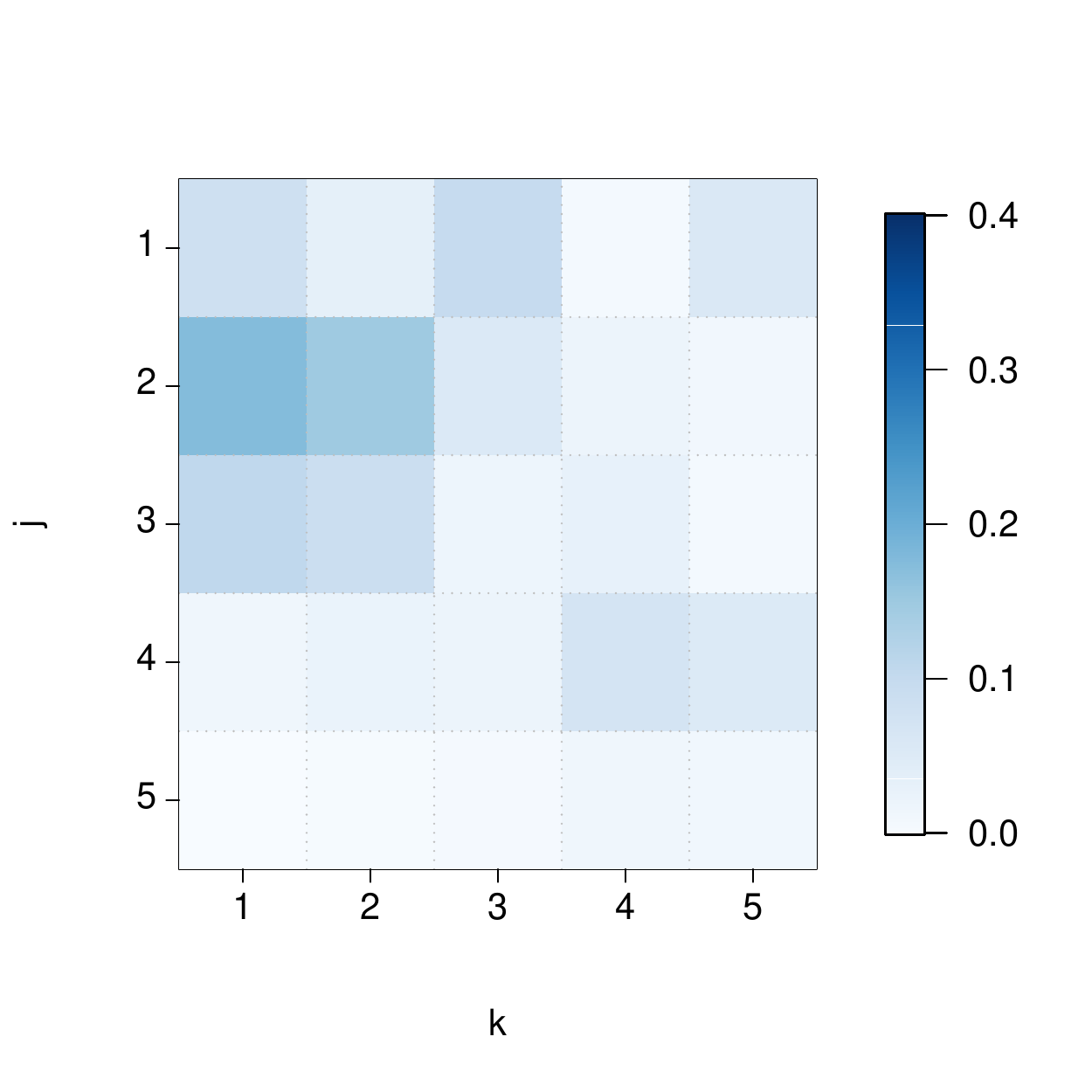}
\end{minipage} %
\begin{minipage}[b]{0.35\linewidth}
        \centering  SPL
        \includegraphics[width=.99\linewidth, clip, trim=0 17 10 55]{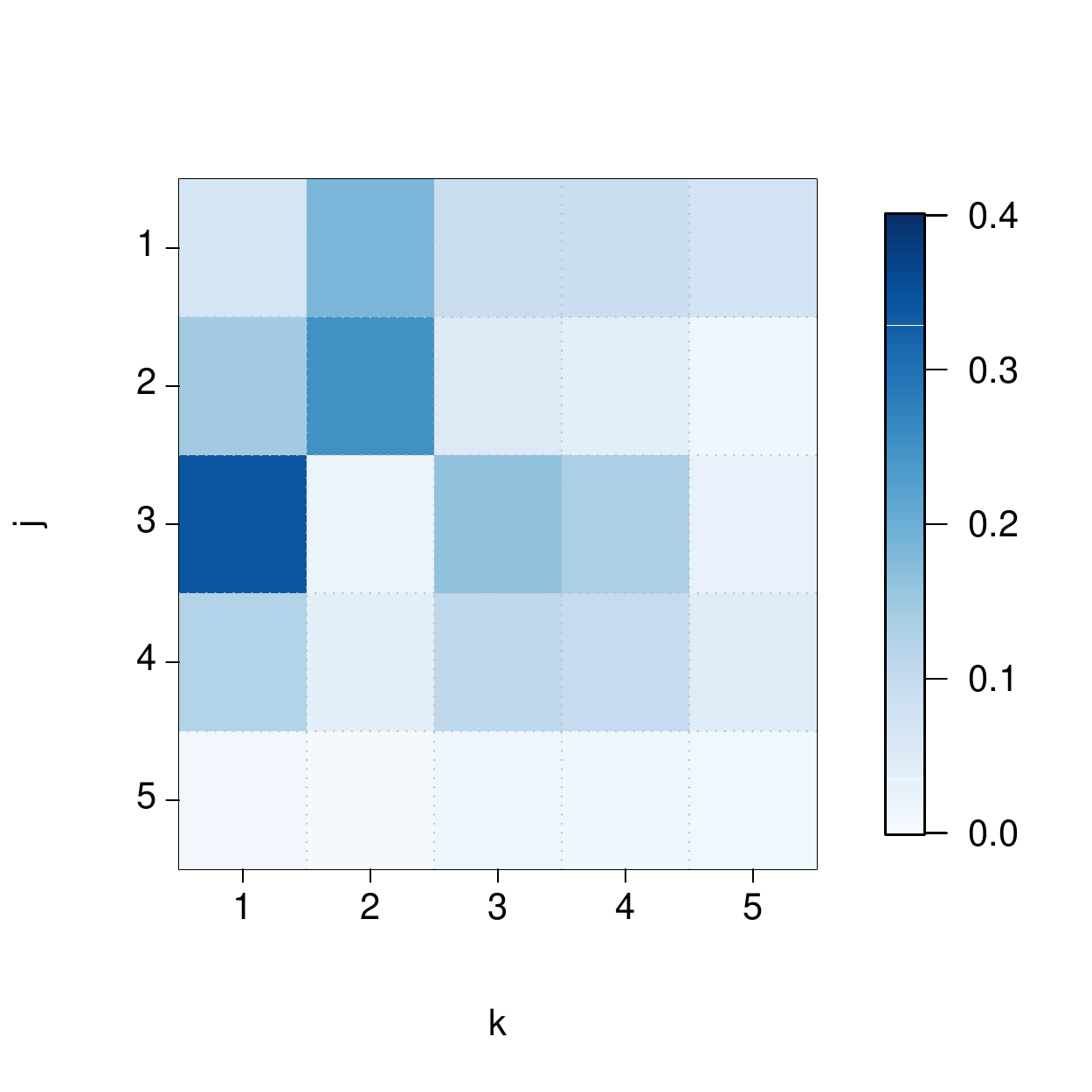}
\end{minipage}
\caption{Song data: heat maps showing $d_{jk} = |\Pr(\tilde{x}_j=k|\D) - \Pr(x_j=k)|$  for the EPL analysis (left) and the SPL analysis (right).
Probabilities based on $L=10$ draws from the predictive distribution per MCMC iteration.}
\label{fig:song_pred_prob_discrepency_images}
\end{center}
\end{figure}

Investigation of the posterior distribution reveals there is no support for the standard (or reverse) Plackett-Luce model(s) with $\Pr(\sigmavec = (3,2,1,4,5)|\D) = 0.9918$, $\Pr(\sigmavec = (5,4,1,2,3)|\D) = 0.0080$ and the remaining posterior mass ($0.0002$) assigned to $\sigmavec=(2,3,1,4,5)$.
It is interesting to see that, although it receives relatively little posterior support, the 2nd most likely choice order parameter value is that given by reversing the elements of the posterior modal value.
It is also worth noting that the posterior modal choice order
$(\sigmavec = (3,2,1,4,5))$ is not contained within the restricted set
considered by \cite{mollica2018algorithms}; this perhaps explains
their conclusion that the (constrained) extended Plackett-Luce model
performs poorly for these data. 
With this in mind we now also question whether the additional complexity of the Extended Plackett-Luce model allows us to better describe the data.
\begin{figure}[b]
\begin{center}
\begin{minipage}[b]{0.49\linewidth}
        \centering 
 	    \includegraphics[width=.99\linewidth, clip, trim=0 0 0 0]{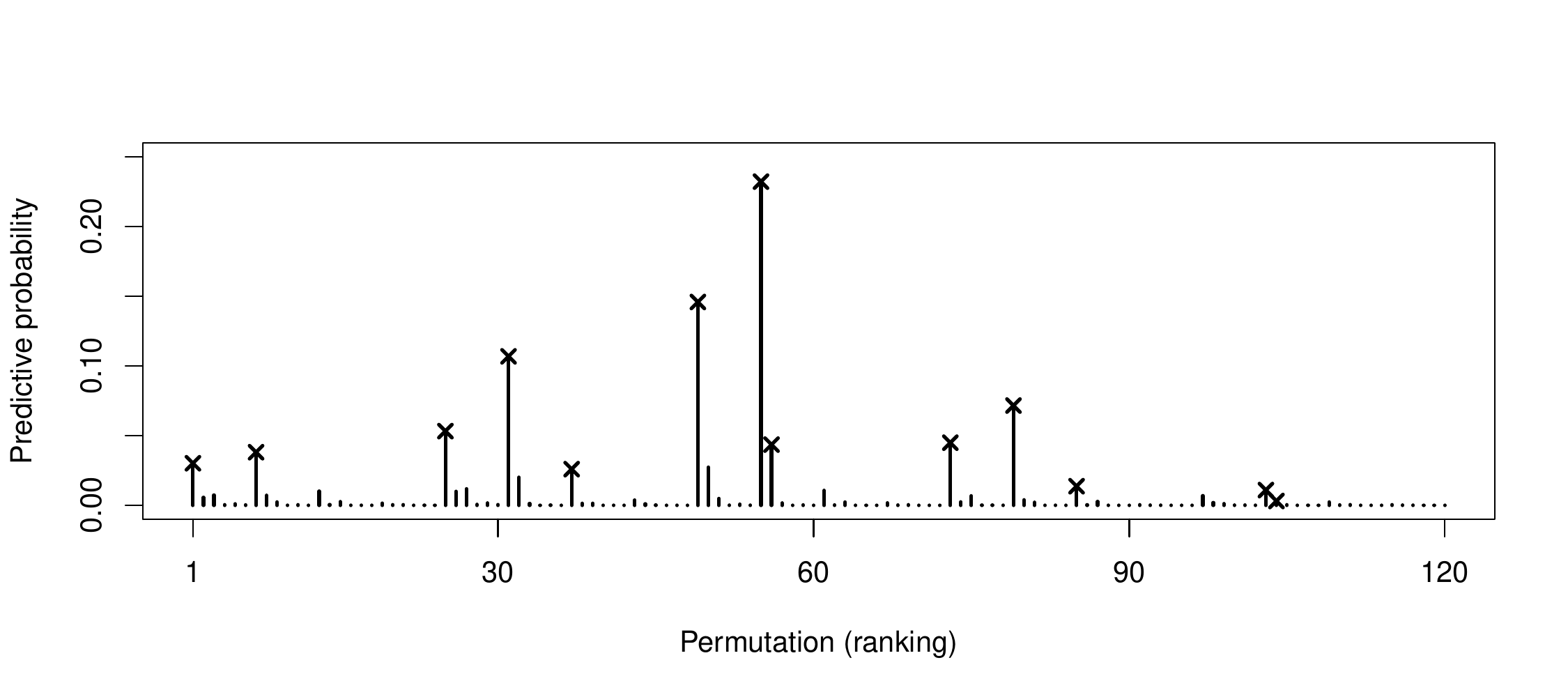}
\end{minipage} %
\begin{minipage}[b]{0.49\linewidth}
        \centering
        \includegraphics[width=.99\linewidth, clip, trim=0 0 0 0]{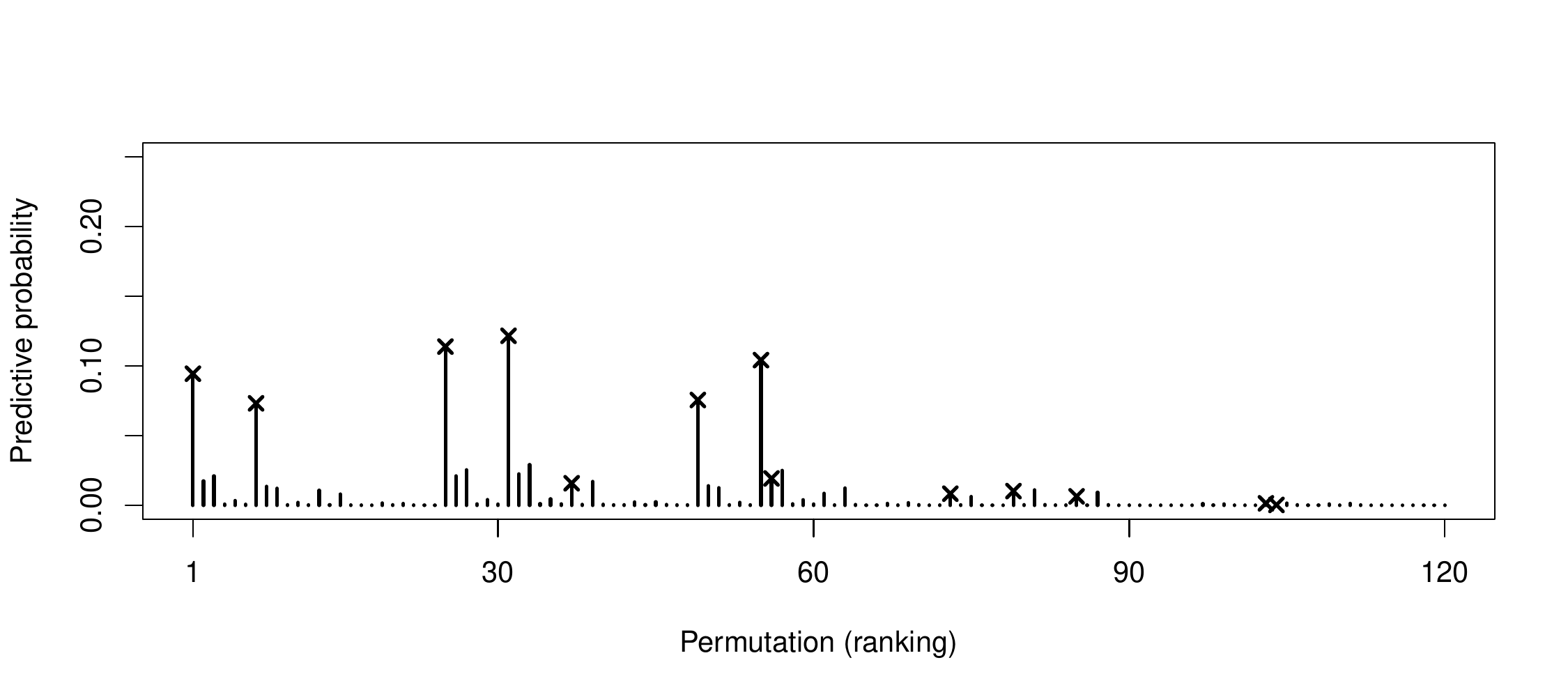}
\end{minipage}
\caption{Song data: Full posterior predictive distribution for each of the $5! = 120$ possible observations $\tilde{\xvec} \in \mathcal{S}_K$ under the EPL (left) and SPL (right) analyses. Crosses ($\times$) highlight the probabilities that correspond to observations within the dataset, that is, those $\tilde{\xvec} \in \D$.}
\label{fig:song_full_post_pred}
\end{center}
\end{figure}
Put another way, does the EPL model give rise to improved model fit.
To this extent we investigate the (full) posterior predictive distribution; where the predictive probabilities for each possible future observation $\tilde{\xvec} \in \mathcal{S}_K$ are computed from the MCMC draws as described in Section~\ref{sec:post_pred}.
For comparative purposes we also compute the predictive distribution obtained from under a standard Plackett-Luce analysis of these data; Figure~\ref{fig:song_full_post_pred} shows the posterior predictive distribution under the extended (left) and standard (right) Plackett-Luce analyses.
The crosses ($\times$) highlight the probabilities that correspond to observations within the dataset (those $\tilde{\xvec} \in \D$), and visual inspection clearly suggests that the observed data look more plausible under the EPL when compared to the SPL.
To further support this conclusion we consider the discrepancies $d_{jk} = |\Pr(\tilde{x}_j=k|\D) - \Pr(x_j=k)|$ as described in Section~\ref{sec:gof}; Figure~\ref{fig:song_pred_prob_discrepency_images} shows these values as a heat map for $j,k \in \{1,\dots,K \}$.
Note that the predictive probabilities $\Pr(\tilde{x}_j=k|\D)$ are computed based on synthetic data simulated from the predictive distribution as discussed in Section~\ref{sec:post_pred} with $L=10$. 
Again these figures suggest the EPL model describes the data much better than the standard Plackett-Luce model.
In particular, there is a rather large discrepancy (0.34) between the predictive and empirical probabilities that entity~$k=1$ (Score) is ranked in position $j=3$ under the SPL analysis.

\begin{figure}[t]
\begin{center}
\begin{minipage}[b]{0.35\linewidth}
        \centering EPL
 	    \includegraphics[width=.99\linewidth, clip, trim= 0 17 10 55]{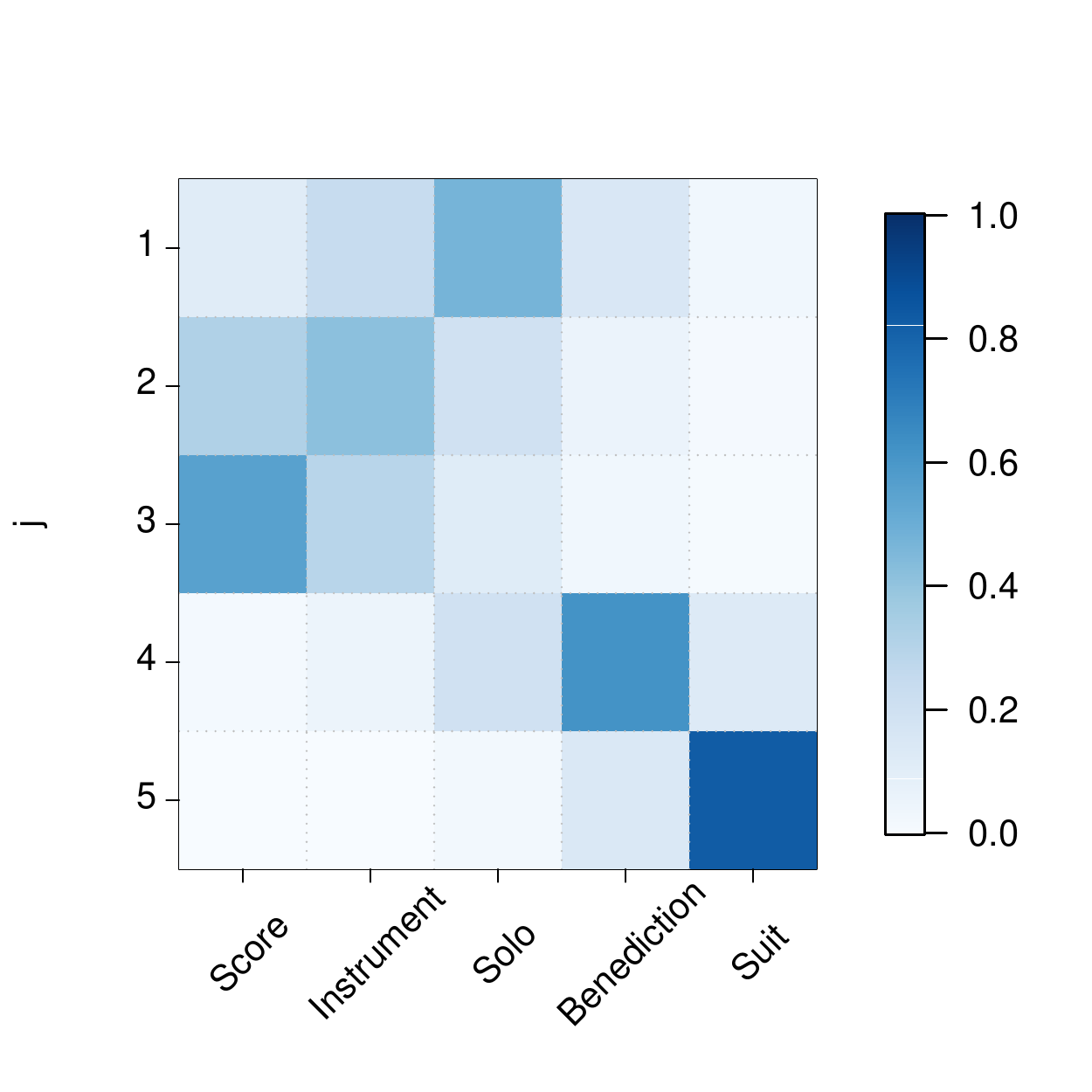}
\end{minipage} %
\begin{minipage}[b]{0.35\linewidth}
        \centering  SPL
        \includegraphics[width=.99\linewidth, clip, trim=0 17 10 55]{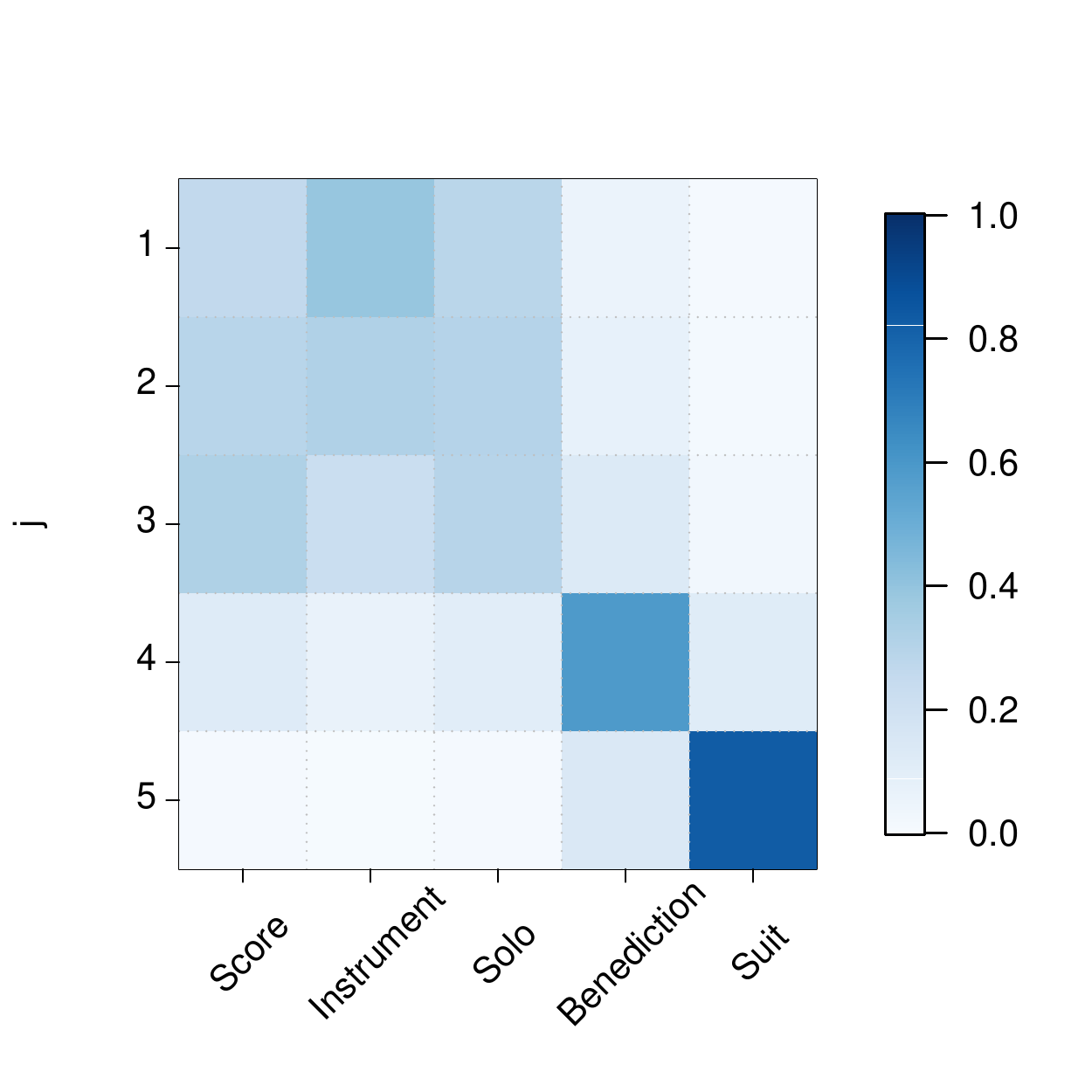}
\end{minipage}
\caption{Song data: heat maps showing $\Pr(\tilde{x}_j=k|\D)$ for the EPL analysis (left) and the SPL analysis (right).
Probabilities based on $L=10$ draws from the predictive distribution per MCMC iteration.}
\label{fig:song_pred_prob_images}
\end{center}
\end{figure}

Turning now to inference (for observable quantities) we again appeal to the posterior predictive distribution.
More specifically we can now use the (predictive) probabilities~$\Pr(\tilde{x}_j=k|\D)$ to deduce the likely positions of entities within rankings.
Figure~\ref{fig:song_pred_prob_images} shows these probabilities as a heat map for $j,k \in \{1,\dots,K \}$.
Focusing on the Extended Plackett-Luce analysis, it is fairly clear that ``Suit'' (5) is the least preferred entity and ``Benediction'' (4) is the 4th most preferred, with relatively little (predictive) support for any other entities in these positions.
There is perhaps more uncertainty on those entities that are ranked within positions $j=1,2,3$, although the figure would suggest that the preference of the entities is (Solo, Instrument, Score, Benediction, Suit).
Indeed this is the modal predictive ranking and has predictive probability~0.232.
Interestingly there appears to be much more uncertainty, particularly for the top~3 entities, under the SPL analysis; further the modal (predictive) ranking is (Instrument, Solo, Score, Benediction, Suit) and occurs within probability~0.122.

\subsection{F1 data}
\label{sec:f1}

We now analyse a dataset containing the finishing orders of drivers within the 2018/19
Formula~1 (F1) season and so we have $n=21$ rank orderings of the $K=20$ drivers.
It will be interesting to see whether we are able to gain more valuable insights using the EPL model when compared to the standard PL model.
In particular  whether we are able to gain any information about the choice order parameter~$\sigmavec$ in this setting as~$K$ is fairly large, relative to~$n$.
The rank orderings considered here were collected from \url{www.espn.co.uk} and also reproduced in the supplementary materials.

Numerous variants of the Plackett-Luce model have previously been developed for the analysis of F1 finishing orders; see \cite{henderson2017comparison} and the discussion therein.
In general, models derived from the reverse Plackett-Luce (RPL) model
appear to perform better than the standard Plackett-Luce model in the sense that they give rise to better model fit.
We choose to incorporate this prior information by letting $\vec{q} = (1,\dots,K)$ and so \textit{a priori} the modal choice ordering is $\hat\sigmavec = (K,\dots,1)$, that is, the choice ordering corresponding to the reverse Plackett-Luce model.
We also take $a_k=1$ (for $k=1,\dots,K$) and so, although we provide information about the likely choice ordering, each rank ordering remains equally likely under this prior specification.
For completeness we also consider an analysis with $\vec{q} = \vec{a} = \vec{1}$ and note that the posterior distribution is not particularly sensative to this choice.
Put another way, these data are rather informative about the choice order parameter~$\sigmavec$ which is perhaps unsurprising given what we have seen from the simulation studies in Section~\ref{sec:sim_studies}.
The following results are based on a typical run of our (appropriately tuned) MC$^3$ scheme initialised from the prior, with appropriate burn-in and thin to obtain 10K (almost) un-autocorrelated realisations from the posterior distribution.
Again we check that $\pi(\sigmavec|\D)$ is consistent under multiple
runs of our algorithm and also use standard MCMC diagnostics on
the~$\lambda$ parameters and the (log) observed data
likelihood~\eqref{eqn:epl_likelihood}. 
This analysis takes around 21 minutes using C code on five threads of an Intel Core i7-4790S CPU (3.20GHz clock speed).

\begin{figure}[b!]
\begin{center}
\begin{minipage}[b]{0.32\linewidth}
        \centering $\vec{q} = (1\dots,K)$
 	    \includegraphics[width=.99\linewidth, clip, trim= 0 17 10 55]{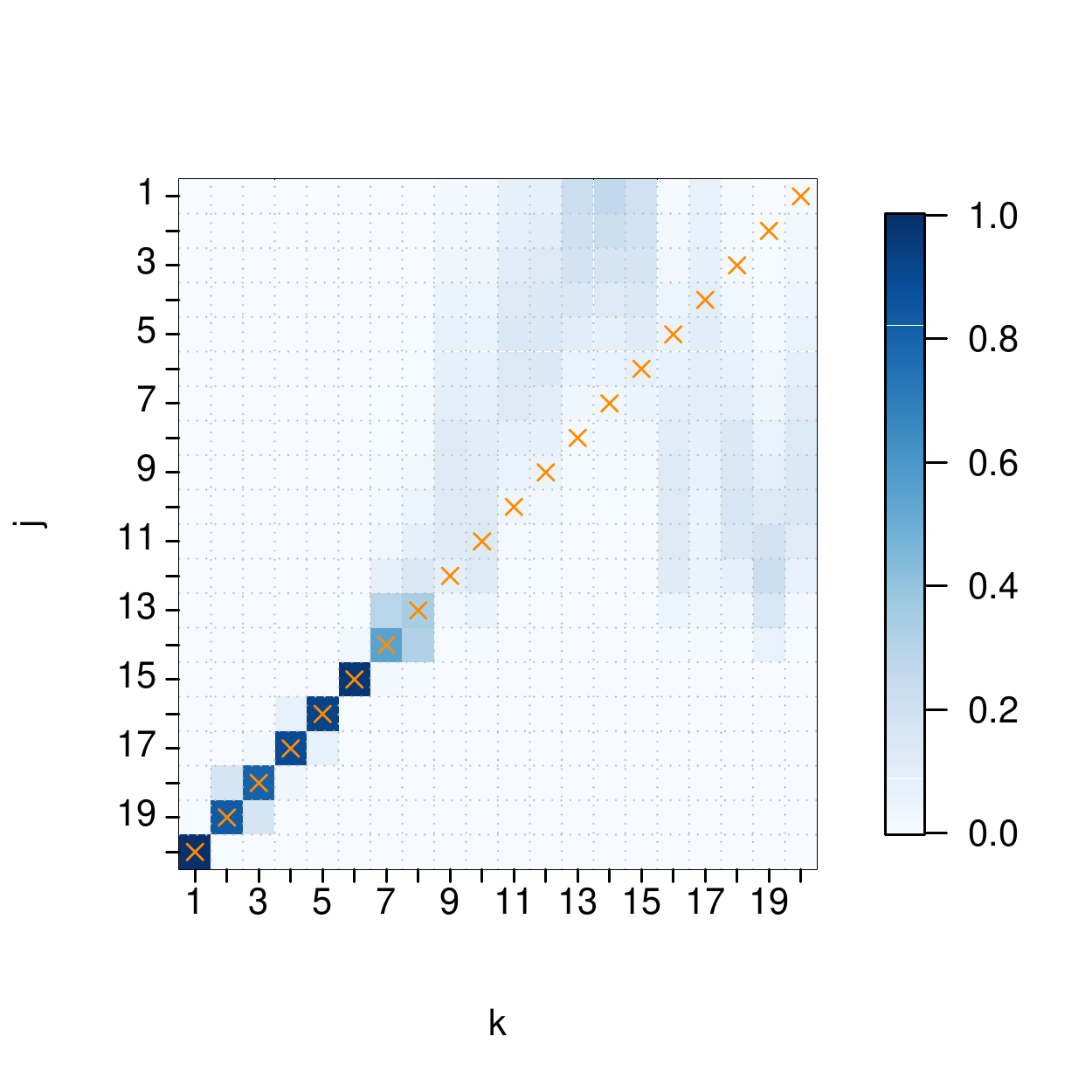}
\end{minipage} %
\begin{minipage}[b]{0.32\linewidth}
        \centering  $\vec{q} = (1\dots,1)$
        \includegraphics[width=.99\linewidth, clip, trim=0 17 10 55]{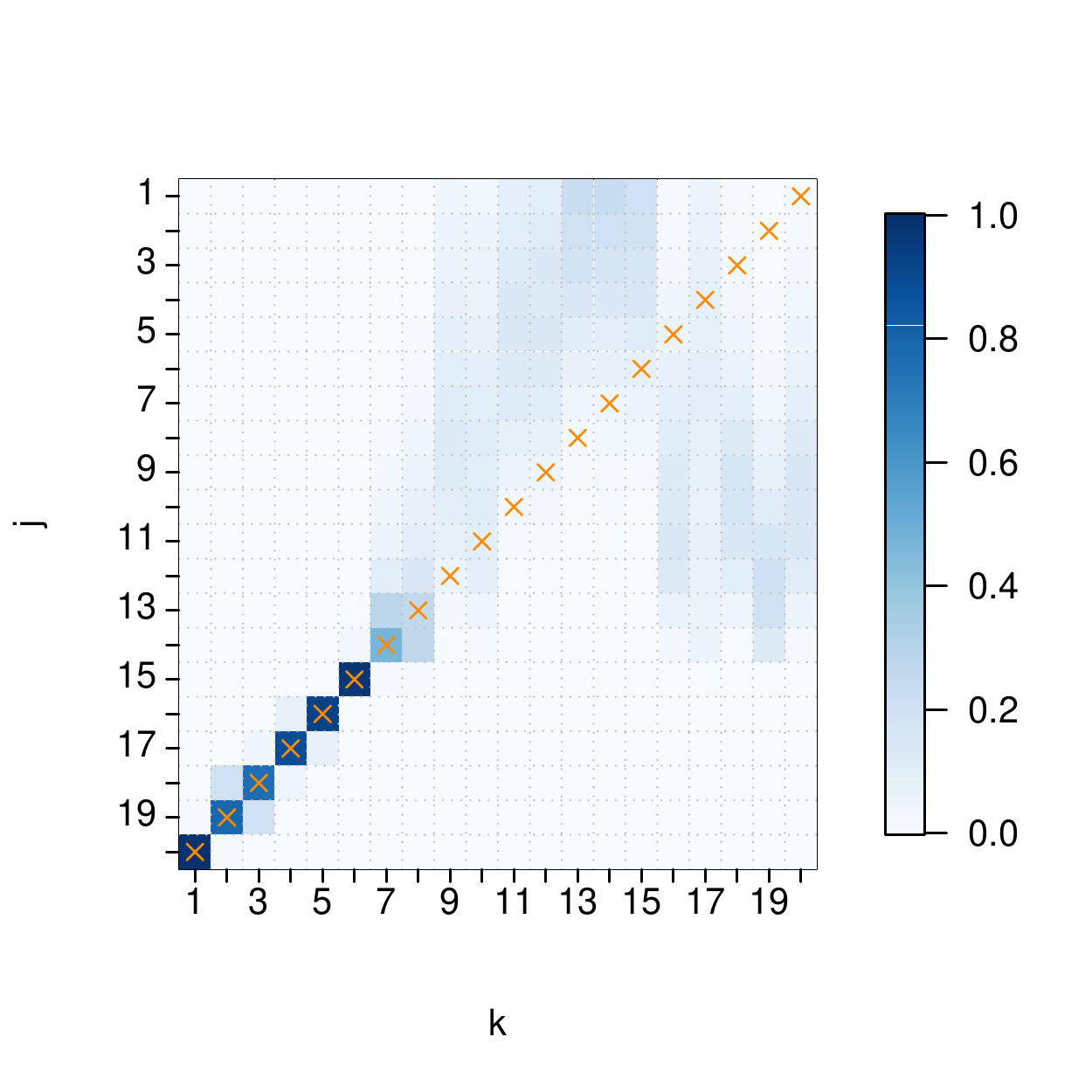}
\end{minipage}
\begin{minipage}[b]{0.32\linewidth}
        \centering  Absolute diff
        \includegraphics[width=.99\linewidth, clip, trim=0 17 10 55]{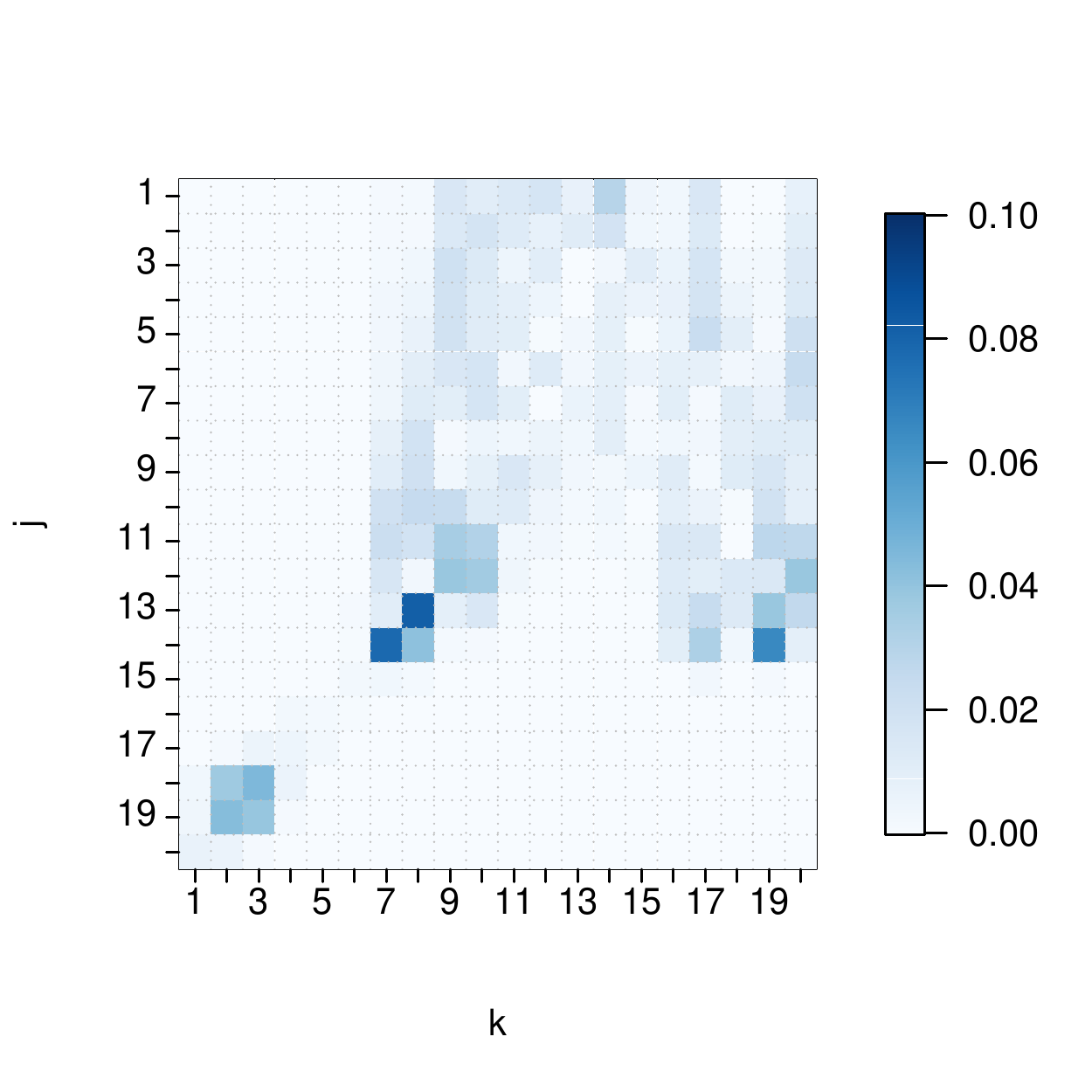}
\end{minipage}
\caption{F1 data: heat maps of $\Pr(\sigma_j=k|\D)$ for the analysis with $\vec{q}=(K,\dots,1)$ (left), $\vec{q}=\vec{1}$ (middle) and the absolute value of the difference (right) for $j,k \in \{1,\dots,K\}$. Crosses ($\times$) highlight the probabilities corresponding to $\sigmavec = (K,\dots,1)$, the reverse Plackett-Luce model.}
\label{fig:f1_sigma_images_paper}
\end{center}
\end{figure}
Investigation of the posterior distribution reveals that there is a large amount of uncertainty on the choice order parameter~$\sigmavec$ and also potential bi-modality within certain ranking stages.
That said, further inspection of the marginal posterior distributions given by $\Pr(\sigma_j=k|\D)$ reveals that there is a surprisingly small amount of uncertainty on the ranks allocated in the 13th-20th stages; see Figure~\ref{fig:f1_sigma_images_paper} (left).
Further within these positions ($\sigma_{13},\dots,\sigma_{20}$) the
ranks allocated are consistent with the choice order parameter
corresponding to the reverse Plackett-Luce model which suggests why
previous authors may have found the RPL model to be preferable to the
SPL model for modelling F1 results.
We also note that these marginal posterior distributions seem fairly robust to the choice of~$\vec{q}$; Figure~\ref{fig:f1_sigma_images_paper} (right) shows the (absolute value of the) discrepancy between the posterior probabilities under each prior choice.

To asses whether the EPL model allows for a good description of these data we again appeal to the posterior predictive distribution.
Here complete enumeration of the posterior predictive probabilities for each $\tilde{\vec{x}} \in \mathcal{S}_K$ is computationally infeasible  as~$K!$ is of $\mathcal{O}(10^{18})$.
We therefore consider the number of times we would expect each of the
top 6 drivers to win a race, feature on the podium (top 3), and also
obtain a points (top 10) finish based on the predictive probabilities
under the EPL model (with $\vec{q}=(K,\dots,1)$) and under an SPL
analysis for comparison.
More specifically Table~\ref{fig:f1_pred_prob_table} shows $n \times \sum_{k=1}^p\Pr(\tilde{x}_j=k|\D)$ for $p=1,3,10$ along with the observed number of times computed from those $\vec{x} \in \D$.
\begin{table}[t]
\setlength\tabcolsep{4pt}
\footnotesize 
\begin{center}
\begin{tabular}{l|ccc||ccc||ccc}
\multicolumn{1}{c}{ } & \multicolumn{3}{c}{Observed} & \multicolumn{3}{c}{EPL}& \multicolumn{3}{c}{SPL} \\
Driver Name (Country) & Wins & Podiums & Points & Wins & Podiums & Points& Wins & Podiums & Points\\
\hline
Lewis Hamilton (GBR) & 11 & 17 & 20 & 10.37 & 16.49 & 20.19 & 5.20 & 12.59 & 20.63\\
Sebastian Vettel (GER) & 5 & 12 & 20 & 4.27 & 12.60 & 19.39 & 3.03 & 8.64 & 19.47\\
Kimi R\"{a}ikk\"{o}nen (FIN) & 1 & 12 & 17 & 1.89 & 8.99 & 18.45 & 1.31 & 4.23 & 14.98\\
Max Verstappen (NED) & 2 & 11 & 17 & 2.21 & 9.70 & 18.65 & 1.35 & 4.32 & 15.00\\
Valtteri Bottas (FIN) & 0 & 8 & 19 & 1.64 & 8.36 & 18.30 & 2.29 & 6.89 & 18.33\\
Daniel Ricciardo (AUS) & 2 & 2 & 13 & 0.59 & 4.90 & 16.79 & 0.76 & 2.57 & 10.86\\
\hline
\end{tabular}
\caption{F1 data: Observed number of wins, podiums (top 3) and
  points (top 10) finishes and also the expected numbers under the
  predictive distributions for the EPL and SPL analyses for the top
  six drivers in the 2018/19 season.}
\label{fig:f1_pred_prob_table}
\end{center}
\end{table} 
Note that the predictive probabilities $\Pr(\tilde{x}_j=k|\D)$ are computed based on synthetic data simulated from the predictive distribution as discussed in Section~\ref{sec:post_pred} with $L=10$. It is interesting to see that the expected number of points (top 10) finishes under both the extended and standard Plackett-Luce models are fairly consistent with the observed data.
However, the shortcomings of the more simple standard Plackett-Luce model become clear if we instead consider the expected number of wins/podiums.
For example we observed that Hamilton won~11 races and the SPL model
would suggest that he would expect to win around 5 races within an F1 season whereas the EPL model suggests 10 wins which is much more consistent with the observed data.
Again additional insight into the question of model fit can be obtained via heat maps showing the discrepancies $d_{jk} = |\Pr(\tilde{x}_j=k|\D) - \Pr(x_j=k)|$  for $j,k \in \{1,\dots,K\}$; these are provided in Section~\ref{sec:f1_supp} of the supplementary materials.

In this setting (large~$K$) we do not have access to the full posterior posterior predictive distribution and so we use an efficient optimisation algorithm based on cyclic coordinate ascent \citep{johnson2018revealing} to find the (global) mode of this distribution.
\begin{figure}[b!]
\begin{center}
\begin{minipage}[b]{0.45\linewidth}
        \centering $\vec{q} = (1\dots,K)$
        \includegraphics[width=.99\linewidth, clip, trim=0 17 10 55]{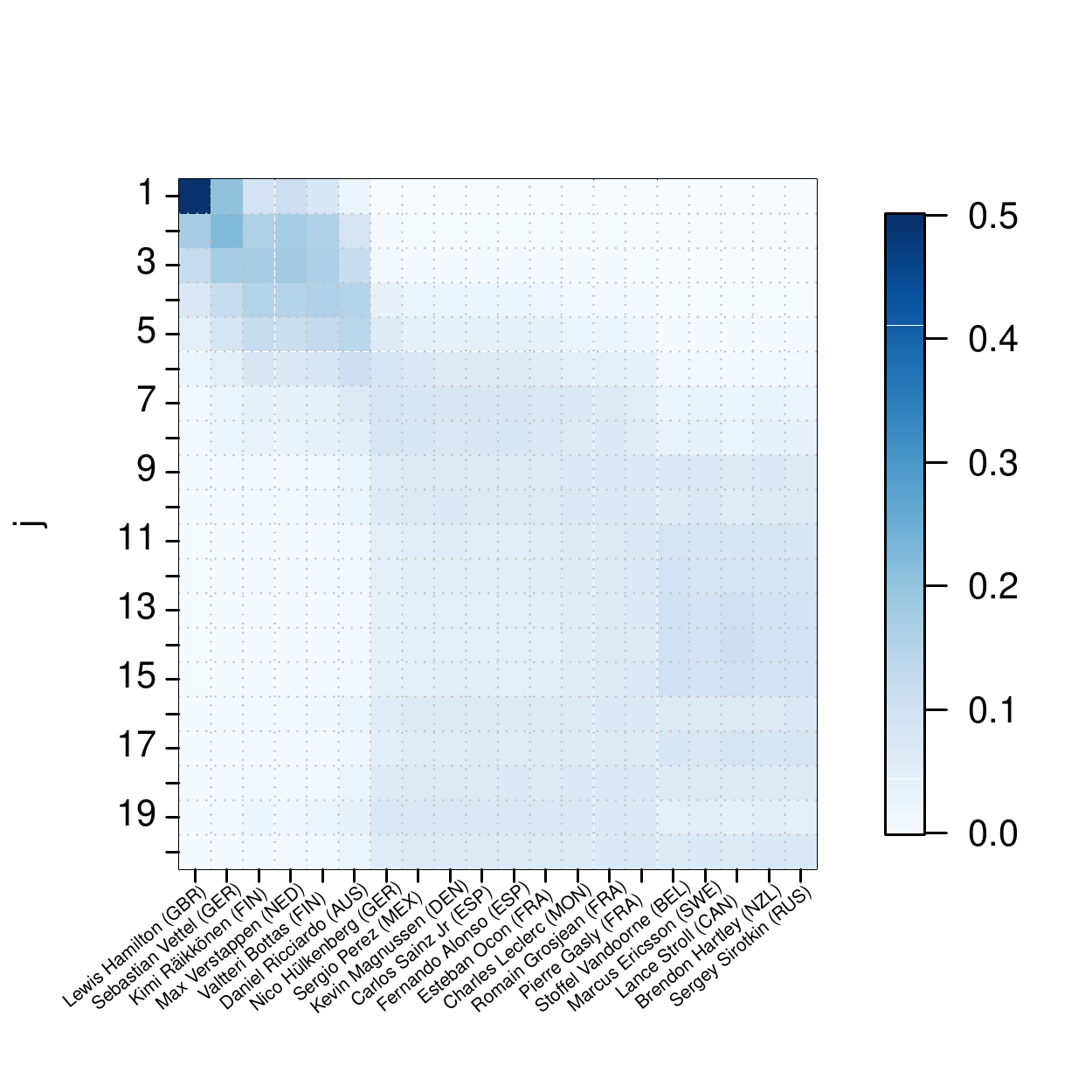}
\end{minipage}
\begin{minipage}[b]{0.45\linewidth}
        \centering  SPL
        \includegraphics[width=.99\linewidth, clip, trim=0 17 10 55]{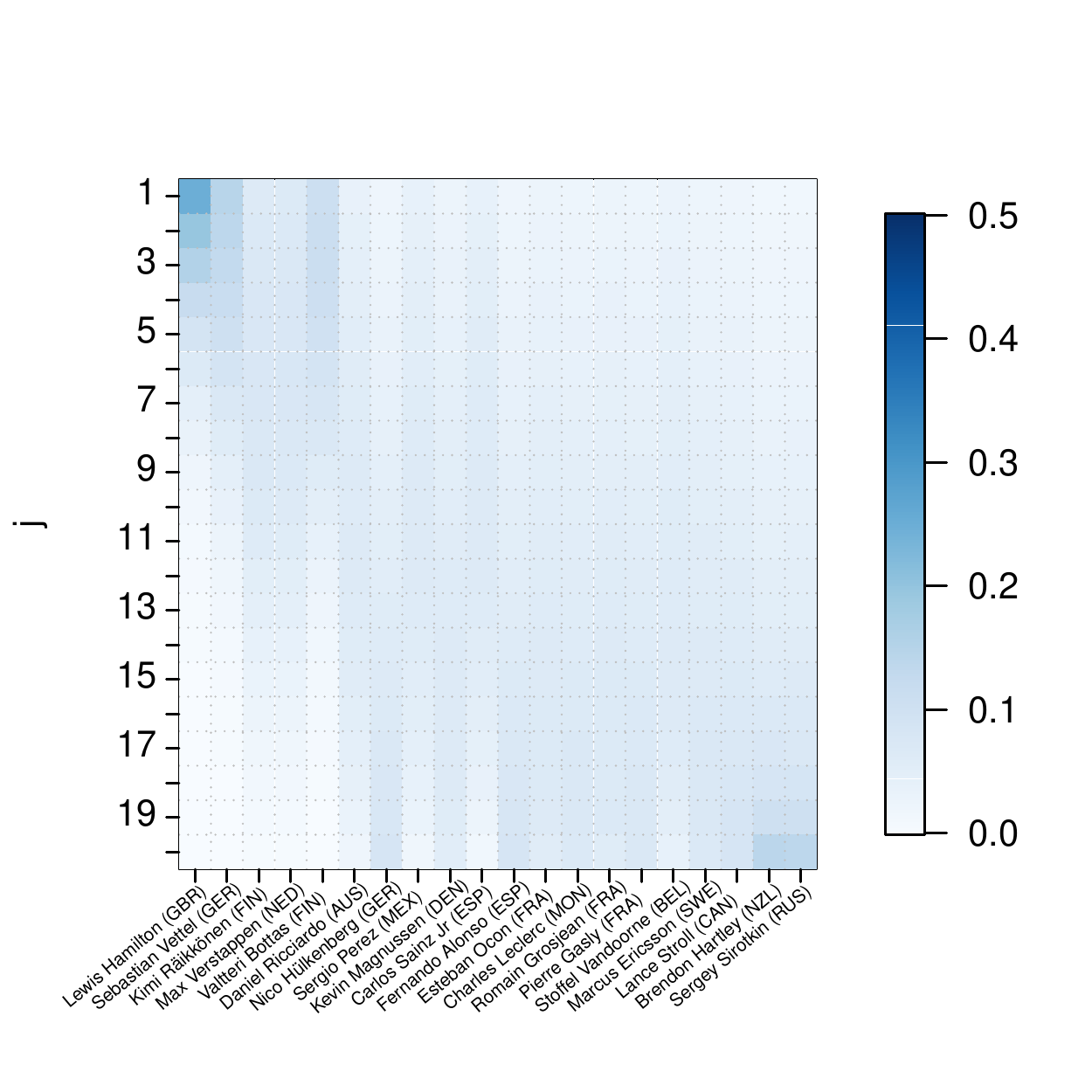}
\end{minipage}
\caption{F1 data: heat maps showing the predictive probabilities $\Pr(\tilde{x}_j=k|\D)$ for the EPL analysis with $\vec{q} = (1\dots,K)$ and SPL analysis for $j,k \in \{1,\dots,K\}$}
\label{fig:f1_pred_prob_images}
\end{center}
\end{figure}
Table~\ref{tab:f1_agg} shows these (aggregate) rankings under both the extended and standard Plackett-Luce analyses along with the observed finishing order based on the driver points (also reported).
\begin{table}[t!]
\footnotesize 
\begin{center}
\begin{tabular}{lcc|cc}
\multicolumn{3}{c}{Final Drivers' Championship standings} \\
Driver Name (Country) & Points &  Rank & EPL & SPL \\
\hline
Lewis Hamilton (GBR) & 408 & 1 & 1 & 1\\
Sebastian Vettel (GER) & 320 & 2 & 2 & 2\\
Kimi R\"{a}kk\"{o}nen (FIN) &251 & 3 & 4 &  5\\
Max Verstappen (NED) &249 & 4 & 3 &  4\\
Valtteri Bottas (FIN) &247& 5 & 5  & 3\\
Daniel Ricciardo (AUS) &170& 6 & 6 & 10\\
Nico H\"{u}lkenberg (GER) &69& 7 & 7 & 6\\
Sergio Perez (MEX) &62& 8 & 11 & 8\\
Kevin Magnussen (DEN) &56& 9 & 12 & 16\\
Carlos Sainz Jr (ESP) &53& 10 & 10 & 9\\
Fernando Alonso (ESP) &50& 11 & 15 & 12\\
Esteban Ocon (FRA) &49& 12 & 17 & 14\\
Charles Leclerc (MON) &39& 13 & 19&17\\
Romain Grosjean (FRA) &37& 14 & 18 &15\\
Pierre Gasly (FRA) &29& 15  & 16 &13\\
Stoffel Vandoorne (BEL) & 12& 16 & 14 &7\\
Marcus Ericsson (SWE) &90 & 17 & 20& 11\\
Lance Stroll (CAN) & 6& 18 & 9 & 18\\
Brendon Hartley (NZL) &4 & 19 & 8 &20\\
Sergey Sirotkin (RUS) &1 & 20 & 13&19\\
\end{tabular}
\caption{F1 dataset: 2018/19 final Drivers' Championship standings
  along with the global mode of the posterior predictive distribution (aggregate ranking) under both the EPL and SPL analyses}
\label{tab:f1_agg}
\end{center}
\end{table} 
It is pleasing to see that both models are able to predict that
Hamilton and Vettel are the two best drivers.
There is some disagreement between those drivers ranked 3rd-5th, however this is perhaps not surprising given these drivers obtained a similar number of points in the 18/19 season.
One of the more concerning observations is that the mode obtained under the SPL model does not contain Ricciardo in 6th place even though he obtained a much larger number of points ($> 100$) than those drivers ranked 7 and below.
For those drivers ranked below 7th there is some general agreement between the ranks under both the extended and standard Plackett-Luce models however we note that there is a large amount of uncertainty about which drivers are placed within these positions; see Figure~\ref{fig:f1_pred_prob_images} which shows the likely position of drivers within the rank orderings based on the predictive probabilities $\Pr(\tilde{x}_j=k|\D)$.

\section{Conclusion}
\label{sec:conc}

We have considered the problem of implementing a fully Bayesian
analysis of rank ordered data using the Extended Plackett-Luce model.
In particular we have considered carefully the problem of prior
specification, proposing a Plackett-Luce model as the prior for the
choice order parameter~$\sigmavec$ and proposing a prior distribution
on the entity parameters that preserves the modal ordering under the
prior predictive distribution. We have also tackled the challenging
issue of posterior sampling of a potentially highly multi-modal
posterior distribution with both discrete and continuous components via
a Metropolis coupled Markov chain Monte Carlo scheme. This has enabled
efficient posterior sampling which potentially facilitates further
analyses based on the Extended Plackett-Luce model and further
extensions of the model. Finally, we have focused on predictive
inference for observable quantities; this admits a natural solution to
the rank aggregation problem and also has facilitated the assessment
of model adequacy.

\section*{Reproducibility}
With reproducibility in mind, the code to run the algorithm outlined in Section~\ref{sec:epl_mc3_alg} can be found at the GitHub repository
\url{https://github.com/srjresearch/ExtendedPL}.
This repository also contains each of the datasets considered within the paper along with detailed comments on how to execute the C code should a user with to perform their own study.
C code for performing a standard Plackett-Luce analysis is also provided.

\section*{Acknowledgements}

This work forms part of the Ph.D.\ dissertation of the first author,
funded by Newcastle University, UK.

\bigskip
A copy of the supplementary materials can be obtained by
contacting  the authors.

\bibliographystyle{apalike}
\bibliography{ref_paper}

\end{document}